\documentclass[linenumbers]{aastex631}
\usepackage{amsmath}
\usepackage{subfigure}
\usepackage{footmisc}

\newcommand{\rvw}[1]{#1}
\newcommand{\revfour}[1]{#1}
\nolinenumbers

\begin{document}

\title{Point Source C-Band Mueller Matrices for the Green Bank Telescope}

\correspondingauthor{Paul Fallon}
\email{paulfallon@telkomsa.net}

\author[0000-0002-4846-1741]{Paul Fallon}
\affiliation{Centre for Space Research, North-West University, Private Bag X1290, 
Potchefstroom 2520, South Africa}

\author[0000-0003-3593-9707]{Derck P Smits}
\affiliation{Department of Mathematical Sciences, University of South Africa, 
Private Bag X6, Florida 1709, South Africa}

\author[0000-0003-4454-2875]{Tapasi Ghosh}
\affiliation{Green Bank Observatory, 155 Observatory Rd, Green Bank, WV 24915, USA}

\author[0000-0002-3168-8659]{Christopher J. Salter}
\affiliation{Green Bank Observatory, 155 Observatory Rd, Green Bank, WV 24915, USA}
\affiliation{Emeritus Scientist, Arecibo Observatory, HC3 Box 53995, Arecibo, PR 00612, USA}

\author[0000-0001-8271-0572]{Pedro Salas}
\affiliation{Green Bank Observatory, 155 Observatory Rd, Green Bank, WV 24915, USA}


\begin{abstract}
C-Band Mueller matrices for the Green Bank Telescope are presented here which 
enable on-sky Stokes parameters for point sources at the beam center to be 
determined. Standard calibrators, 3C138 and 3C286, were observed using the Spider
program to steer the telescope across a broad range of Right Ascensions on both 
sides of the zenith transit. For this analysis, only the observations at the 
peak of the Spider pattern were used rather than the full sweep of the runs. 
Therefore, the results presented here only apply to point sources at the beam 
center. The Mueller matrices are shown to vary with frequency and with use of 
the Hi-Cal or Lo-Cal noise diodes, \revfour{due to the relative calibration 
gain between the $X$ and $Y$ components of the feed.} 

\revfour{However, the relative calibration gain can be determined from 
observations of a source with known polarization. Correcting the data for the 
relative calibration gain prior to data analysis allows for use of a frequency 
independent Mueller matrix. This generic Mueller matrix is shown to provide 
reliable C-Band polarization measurements.}   
\end{abstract}

\section{Introduction} 
Polarization properties of radio signals provide valuable information about 
the source, its surrounding region, and the interstellar medium. The versatile 
GBT astronomical spectrometer \citep[VEGAS,][]{Prestage_2015} on the Robert 
C.~Byrd 100 m Green Bank Telescope (GBT) allows full Stokes measurements to 
be made, but as with all radio telescopes, instrumental effects distort the 
observed signal. The feed, telescope structure, dish surface, coaxial cables, 
optical fibers and electronics can introduce gain and phase differences in 
the individual measurement channels which modify the polarization. Measured 
Stokes values can be corrected using an empirically determined Mueller matrix 
to obtain the source polarization parameters. 

Mueller matrices have been determined previously for the GBT.  The L-band 
system was calibrated by \citet{Liao_et_al_2016} and the C-Band by 
\citet{RH2006}. However, subsequent system changes, e.g.~replacement of the 
C-Band receiver in 2014, result in the calibrations no longer being valid. 

\section{Observation}
The C-Band receiver on the GBT contains a single waveguide located at the 
Gregorian focus of the telescope, feeding into a cooled ortho-mode transducer 
(OMT) that produces two orthogonal linearly polarized outputs. The linear 
$X$ component of the feed is aligned parallel to the elevation axis 
(horizontal) and the $Y$ component is aligned perpendicular to the elevation 
axis (vertical) on the alt-az mounted telecope. Resonances within the bandpass
can arise in the OMT, but none are listed for the C-Band system. There are two 
noise diodes (NDs) with levels of  $\sim10$\% (Lo-Cal) and $\sim100$\% (Hi-Cal) 
of the system temperature that can be used for flux calibration. The nominal 
frequency range of the C-Band system is 3.95 -- 7.8~GHz. Further details 
regarding the receiver system can be found in the GBT Observers Guide 
v4.0\footnote{Available at https://www.gb.nrao.edu/scienceDocs/GBTog.pdf.}.

The VEGAS spectrometer was used for all our observations; it consists of 
eight independent spectrometers (banks) that can be used simultaneously, 
each one producing a full set of polarization products. The auto- ($XX$ and 
$YY$) and cross- ($XY$ and $YX$) correlation terms produced by the VEGAS 
spectrometer \rvw{were calibrated as described in Section \ref{sec:analysis} 
and Appendix \ref{SpectraCal}}, and then used to obtain the observed Stokes 
parameters, which are defined as
\begin{eqnarray}
I_{\text{obs}} & = & XX + YY \label{eqn1}\\
Q_{\text{obs}} & = & XX - YY \label{eqn2}\\
U_{\text{obs}} & = & 2*XY \label{eqn3}\\
V_{\text{obs}} & = & 2*YX \ . \label{eqn4}
\end{eqnarray}

The convention used here is that the polarization angle is measured East of 
North, as specified by IAU Commissions 25 and 40. The Stokes $V$ component 
uses the standard IAU definition for the circular component in radio astronomy 
\citep{IEEE18}, i.e.~$V$ = RCP -- LCP\@. The double reflection produced by the 
main dish and the subreflector ensures that the Stokes $V$ component from the 
correlator matches that of the incoming signal. When comparing results from 
other telescopes, it is important to check what definitions have been used 
for the Stokes parameters \citep{RH2021}. 

The Mueller matrix measurements comprised observing linearly polarized point 
sources over a range of parallactic angles (P.A.s) on both sides of the prime 
meridian. To accomplish this the GBT Spider scan routine was used. It consists 
of four scans centered on-source, starting and ending three beam widths off 
center. The on-sky trajectory of the Spider scan is shown in Figure~D.1 on 
page 213 of the GBT Observers Guide v4.0. The Hi-Cal ND was fired for 10 s 
(with 1~s switching period) before and after each run along a leg of the 
Spider scan. The Spider routine acquired 80 observations of 1 s duration each 
along each leg. Observation 40 was the on-source measurement which had the 
maximum flux reading along each leg. Each full Spider procedure lasted 
$\sim 14$ min.

The Spider run configuration comprised, the eight banks of the spectrometer 
using VEGAS mode 4, each covering a bandwidth of 187.5\,MHz with central 
frequencies evenly distributed from 4.3 to 7.1~GHz (see Table~\ref{table:Mm1}). 
Only data from the middle 1/3 of the bandwidth ($\Delta \nu \sim 60$ MHz) were 
used for the analysis. Two linearly polarized calibration sources were measured. 
3C286 was observed for 3 hr on 2021 January 02 and 3C138 for 6 hr on 2021 
January 13. 

Spectral line frequency-switched observations using VEGAS mode 15 and the 
Lo-Cal ND (with 1~s switching period) were made on 2021 January 05 and 
2021 November 27. The eight banks of the VEGAS spectrometer were centered 
on frequencies of $\nu = 4.765, 4.751, 4.660, 4.830, 6.033, 6.049, 6.181, 
6.668$~GHz each with a bandwidth of 11.72~MHz, but only the middle 1/3 of 
the data was used, giving $\Delta \nu \sim4$ MHz. On 2021 January 05, 
3C138 was observed before our target source for 210 s to calibrate the 
polarization and the source B0529$+$075 for 6 min to calibrate the flux. 
On the November run, 3C138 and B0529$+$075 were observed (for 120 s and 
180 s respectively) using both the above spectral line and the evenly 
spaced Spider run frequency settings before our source observation, and 
3C138 was observed again with both frequency settings after our source 
observations. All observations were part of project \textit{AGBT20B\_424}. 
\\

\section{Analysis}\label{sec:analysis}
Analysis of the auto-correlation terms is straightforward and can be 
accomplished using standard GBTIDL routines such as \textit{getfs} and 
\textit{getps}, whereas the cross-correlation terms need to be reduced 
using polar coordinates. \rvw{Data calibration is achieved using the 
ND-on and ND-off modes of either the Hi-Cal or Lo-Cal ND. The $XX$ and 
$YY$ spectra are each calibrated separately, whereas the $XY$ and $YX$ 
spectra are calibrated simultaneously in polar coordinates as described 
in Appendix \ref{SpectraCal}.}  
\revfour{For these reductions the same $T_\text{cal}$ calibration constant has 
been used for all four components as explained in Section \ref{sec:GBTcal} 
and Appendix \ref{SpectraCal}}.
A new set of GBTIDL analysis routines has 
been developed to perform the reductions \citep{Fallon2022}.
These routines have been rigorously tested \rvw{and shown} to produce the 
same \rvw{calibrated outcomes} as the standard routines for the 
auto-correlation terms. Observed Stokes parameters were determined 
from the calibrated spectra for each spectrometer bank using 
equations (\ref{eqn1}) -- (\ref{eqn4}).  

Because the GBT has an alt-azimuth mount, the P.A.~of the feed rotates 
on the sky. The observed parameters are rotated from the sky frame by the 
rotation matrix $\boldsymbol{M_{\text{sky}}}$ and then distorted further 
by instrumental effects that are described in terms of the Mueller matrix 
$\boldsymbol{M}_\mathrm{Mueller}$. The observed Stokes values are a product 
of the Mueller matrix, the rotation matrix and the source Stokes parameters, 
i.e. 
\begin{align}
\begin{bmatrix}
I_\textrm{obs}\\ Q_\textrm{obs}\\ U_\textrm{obs}\\ V_\textrm{obs}
\end{bmatrix}
= \boldsymbol{M_\textrm{Mueller} \: \cdot } \: \boldsymbol{M_\textrm{sky}}
\begin{bmatrix}
I_\textrm{src}\\ Q_\textrm{src}\\ U_\textrm{src}\\ V_\textrm{src}\
\end{bmatrix}.
\label{eqn5}
\end{align}

Once the appropriate Mueller matrix is determined, source Stokes parameters can 
be obtained from the inverse of equation (\ref{eqn5}),  
\begin{align}
\begin{bmatrix}
	I_\textrm{src}\\ Q_\textrm{src}\\ U_\textrm{src}\\ V_\textrm{src}
\end{bmatrix}
= (\boldsymbol{M_\textrm{Mueller} \: \cdot } \: \boldsymbol{M_\textrm{sky})^{-1}
\begin{bmatrix}
I_\textrm{obs}\\ Q_\textrm{obs}\\ U_\textrm{obs}\\ V_\textrm{obs}
\end{bmatrix}
= (\boldsymbol{M_\textrm{sky}})^{-1} \: \cdot } \: (\boldsymbol{M_\textrm{Mueller}})^{-1}
\begin{bmatrix}
I_\textrm{obs}\\ Q_\textrm{obs}\\ U_\textrm{obs}\\ V_\textrm{obs}
\end{bmatrix}.
\label{eqn6}
\end{align}

The theory used to determine single dish Mueller matrices has been 
described by \citet{Heiles_et_al_2001, Heiles2002, RH2021}. These 
equations are presented in Appendix \ref{MMeq} where it can be seen that 
the observed fractional Stokes parameters are functions of the P.A., the 
fractional source Stokes parameters, and the five Mueller matrix parameters. 
The observed fractional Stokes parameters from the Spider runs were fitted 
to equations (\ref{eqn10}) -- (\ref{eqn12}) using the Add-in Solver in 
Excel\footnote{https://support.microsoft.com/en-us/office/load-the-solver-add-in-in-excel-612926fc-d53b-46b4-872c-e24772f078ca}. 
The starting values for the fitting routine used the values for the Mueller 
matrices presented in \cite{RH2006}. To check that the routine is finding 
an absolute rather than a local minimum, tests were done with a wide range 
of starting values. In all cases the iteration returned the same final 
values. The final values are the five Mueller matrix parameters and the 
source parameters $(Q/I, U/I)_\textrm{src}$. In Appendix \ref{MMeq} it is 
pointed out that we assume $V_\textrm{src} = 0$, however the fitting process
can be conducted without using this constraint.

\rvw{Mueller matrices are derived from our Spider scan observations, which covered 
a wide range of P.A.s. Fewer observations can be used to determine Mueller matrices 
(but this might increase the uncertainties in the values), provided at least 
two observations are made at different P.A.s and the observed fractional 
Stokes values outnumber the number of unknowns in equations (\ref{eqn9}) -- 
(\ref{eqn12}).} Observations of 3C138 in 2021 November were made at two 
P.A.s (before and after our spectral line source observations) using the Lo-Cal ND to 
calibrate the spectra. Fitting equations (\ref{eqn9}) -- (\ref{eqn12}) to 
the observations and using values of $(Q/I, U/I)_\textrm{src}$ determined from 
the Spider runs, provided sufficient constraints for the Mueller matrix parameters 
to be uniquely determined using Solver, but with higher uncertainty than for 
the Spider runs. 

\section{Determining the Mueller matrix}\label{sec:results}
\begin{figure} [t]
	\begin{center}
	\subfigure[]{
       \includegraphics[width=0.49\textwidth, clip, trim=3.3cm 1.8cm 3.4cm 1.8cm]
       {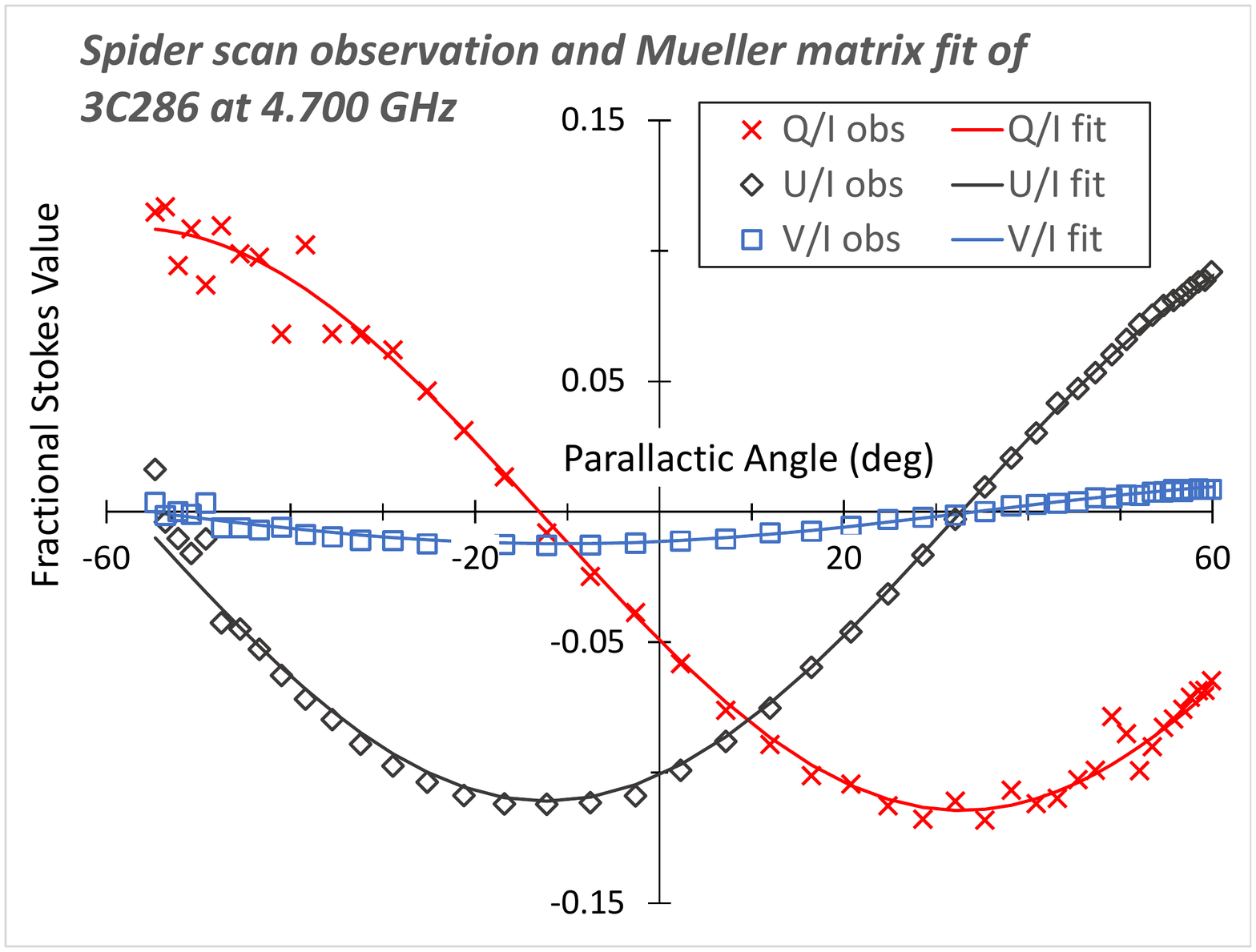}}          
    \subfigure[]{
       \includegraphics[width=0.49\textwidth, clip, trim=2.8cm 1.4cm 2.8cm 1.4cm]
       {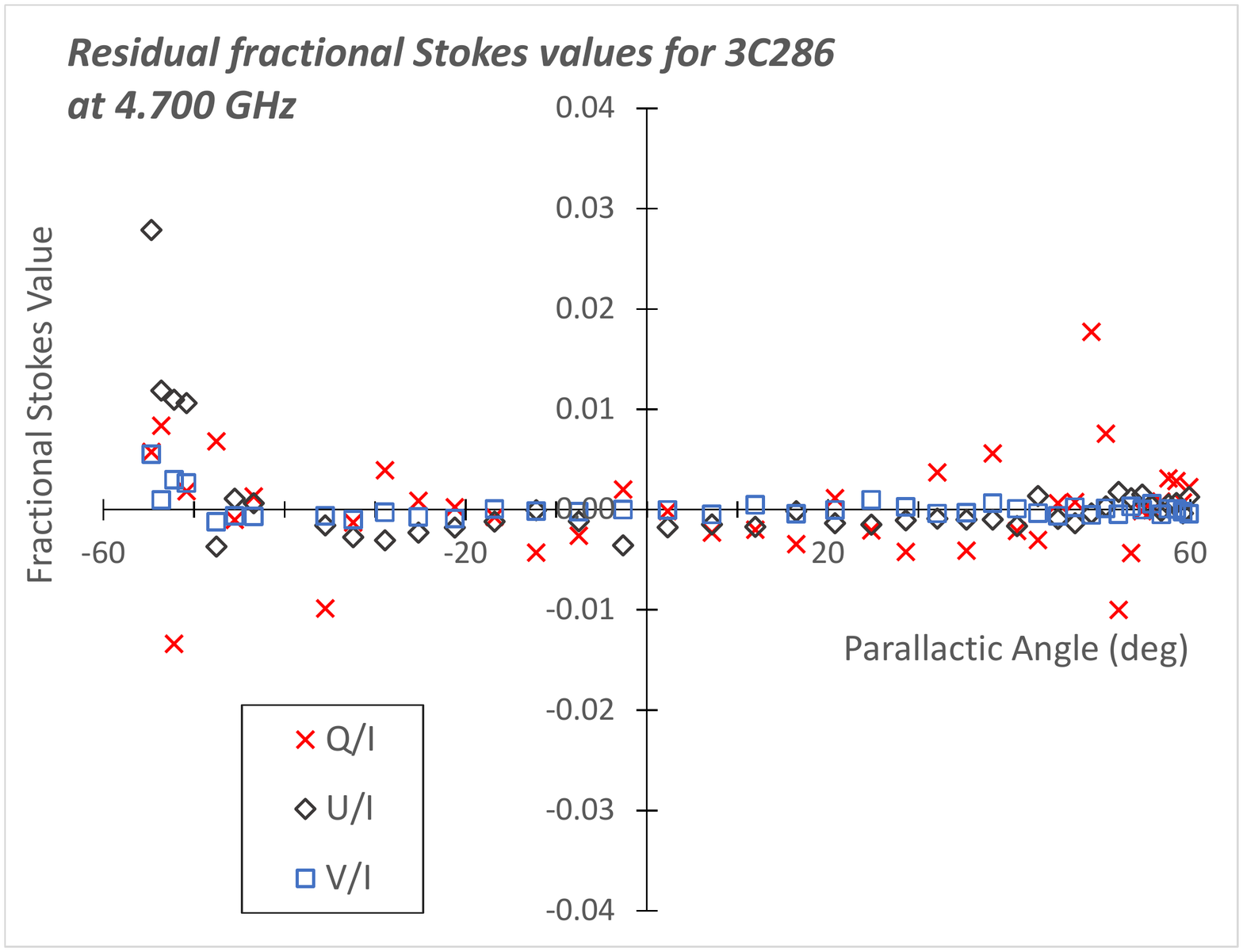}}
    \end{center}
    \caption{(a) Observed 3C286 fractional Stokes parameters as a function 
    of P.A. fitted with Mueller matrix equations. (b)~Differences between 
    the observed and fitted fractional Stokes parameters, indicating no 
    systematic variation. \label{Fit}}
\end{figure}

\begin{figure} [t]
	\begin{center}
	\subfigure[]{
       \includegraphics[width=0.49\textwidth, clip, trim=2.6cm 1.8cm 2.4cm 1.6cm]
       {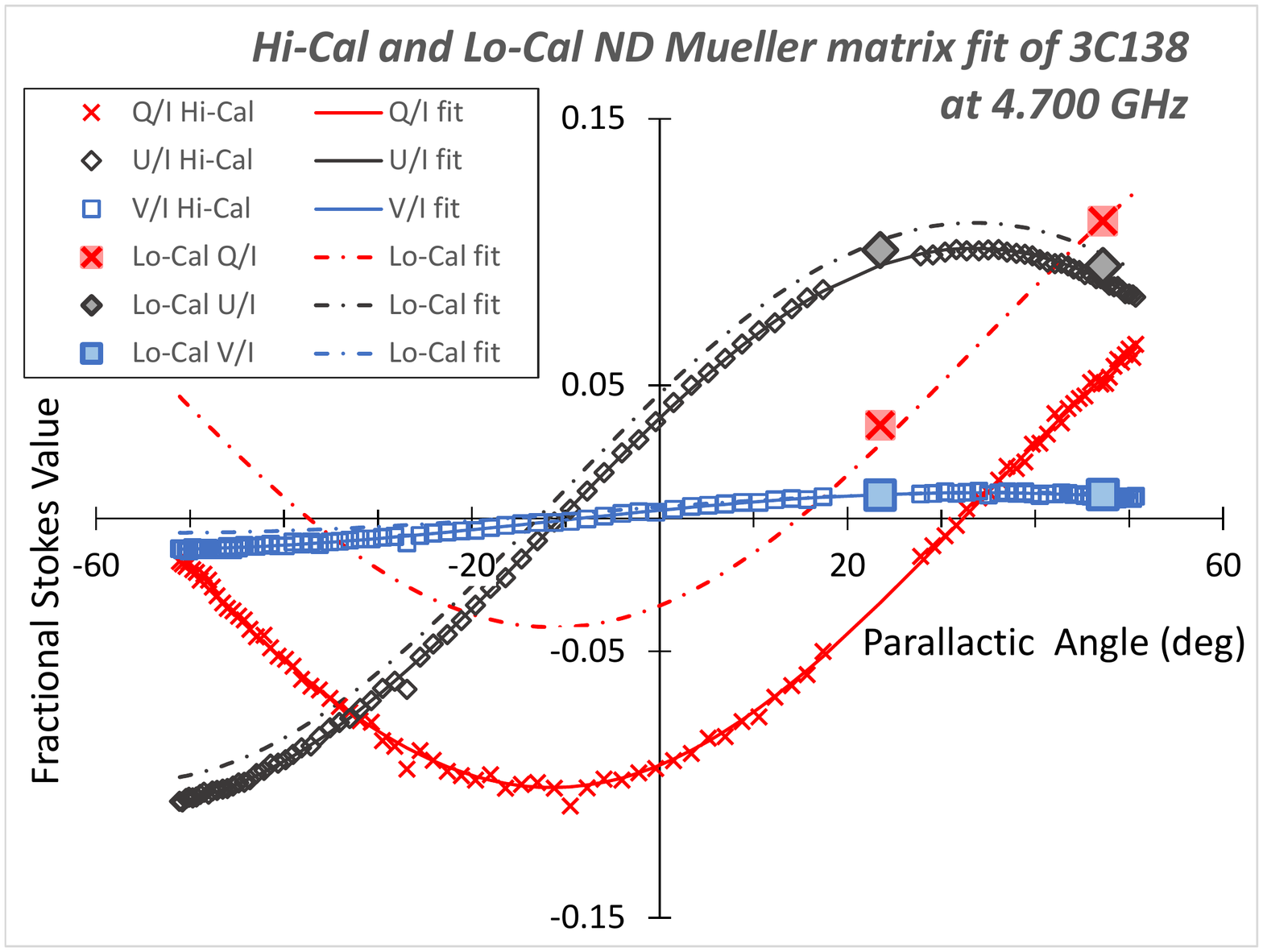}}          
    \subfigure[]{
       \includegraphics[width=0.49\textwidth, clip, trim=3.0cm 1.8cm 3.0cm 1.8cm]
       {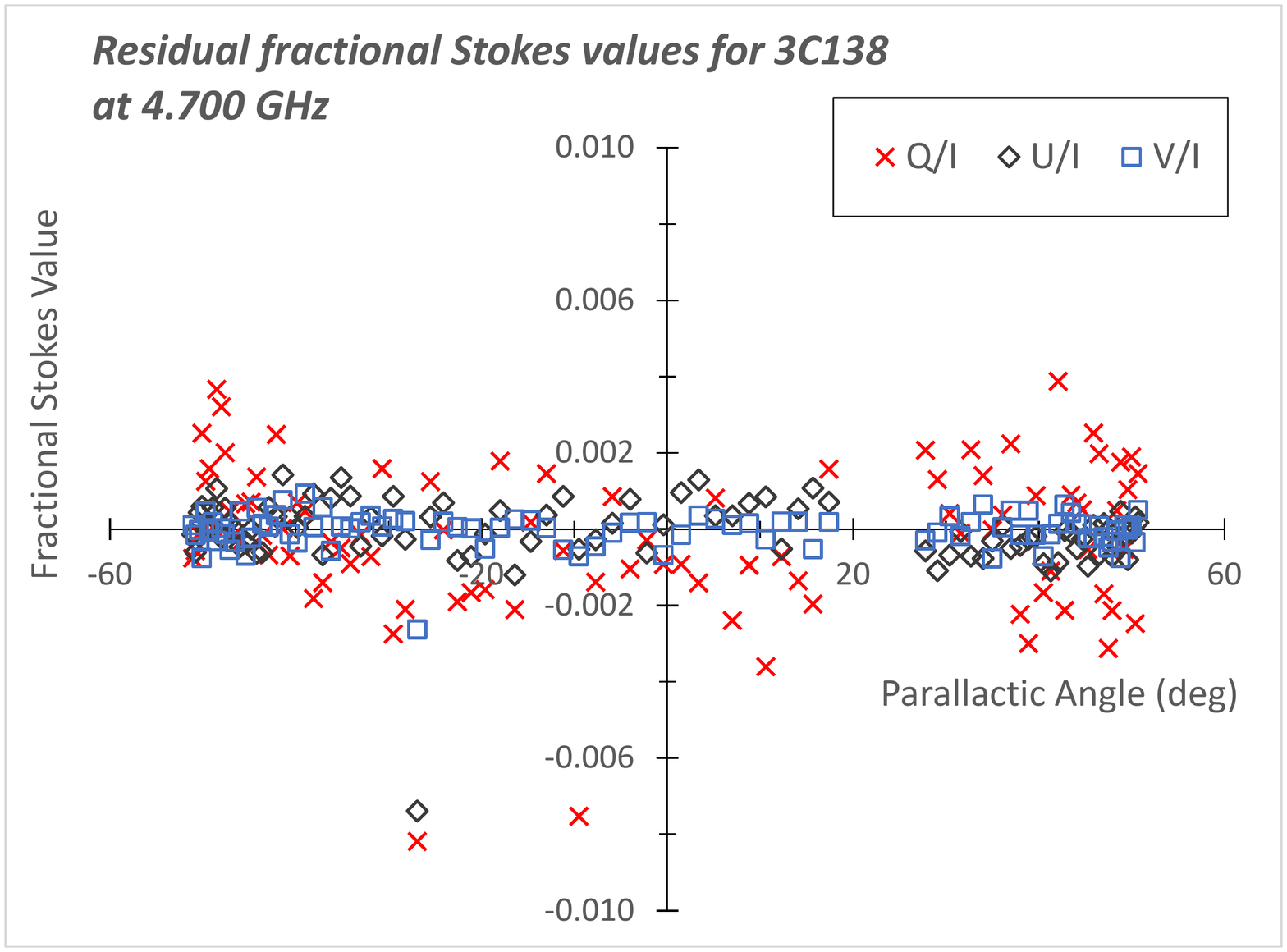}}
    \end{center}	
    \caption{(a) Observed 3C138 fractional Stokes parameters as a function 
    of P.A., fitted with Mueller matrix equations using solid line. The 
    Lo-Cal ND observations are shown with larger symbols and dot-dashed 
    lines, and indicate a difference with the Hi-Cal ND results. Because all 
    the Lo-Cal Q/I, U/I and V/I parameters are fitted simultaneously, 
    the Lo-Cal Q/I and U/I values are offset from the calculated curve.  
    (b)~Differences between the observed and fitted fractional Stokes 
    parameters for the Hi-Cal ND data points, indicating no systematic 
    variation. The vertical scale is different to that in Figure 1(b) 
    which has more scatter. \label{Fit2}}
\end{figure}

The observed fractional Stokes parameters as a function of P.A.~for the Spider 
run on 3C286 are shown in Figure~\ref{Fit} and for 3C138 in Figure~\ref{Fit2}. 
The Mueller and rotation matrix equations fitted to the observed values are 
shown as solid lines in Figures~\ref{Fit} and \ref{Fit2}. The middle C-Band 
frequencies display better quality fits than the outside frequencies; the 
most data scatter was observed in the 7.1~GHz band. 

\begin{figure} [t]
    \begin{center}
    \subfigure[]{
       \includegraphics[width=0.49\textwidth, clip, trim=2.7cm 1.8cm 2.7cm 2.8cm]
       {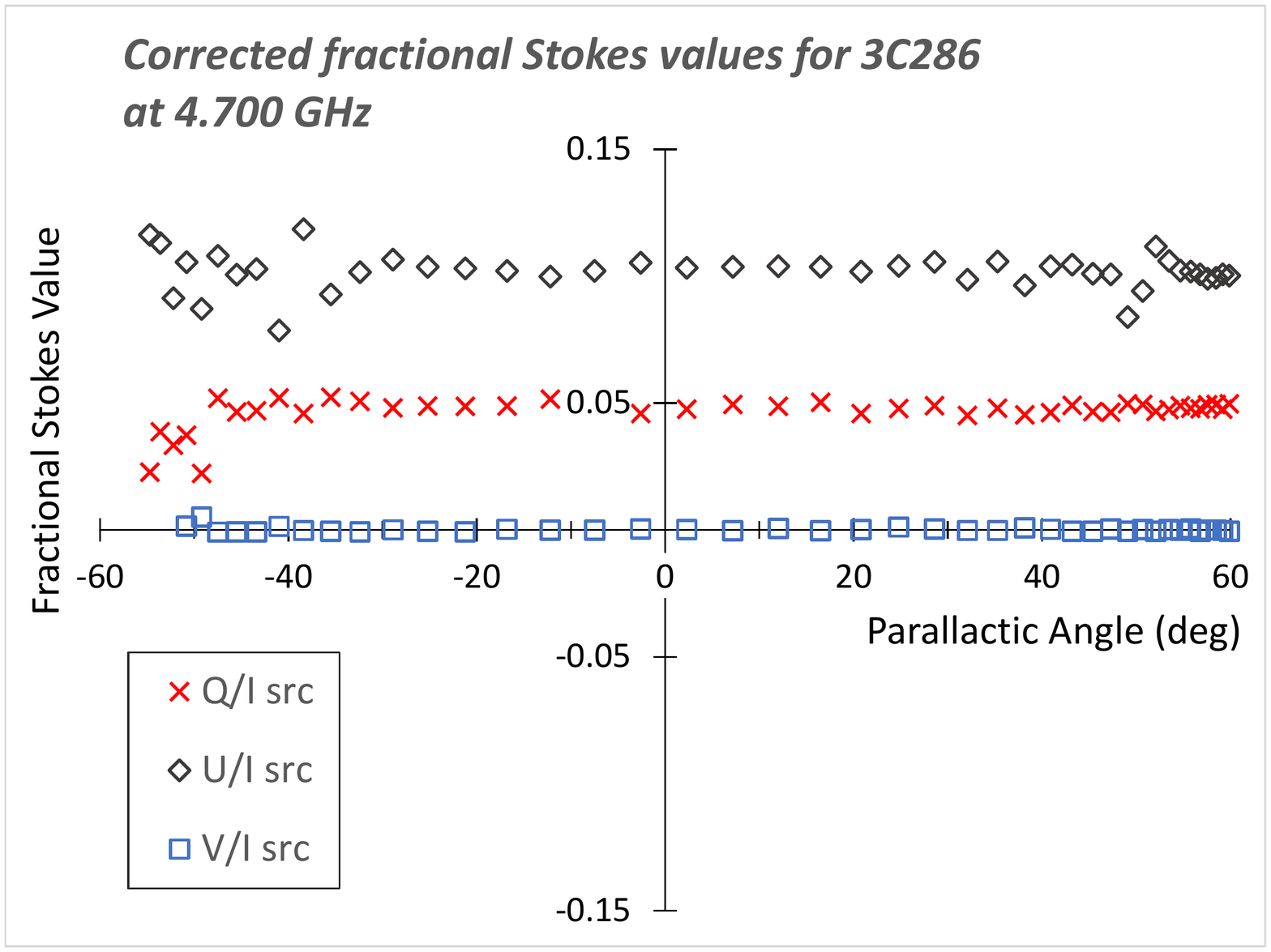}}          
    \subfigure[]{
       \includegraphics[width=0.49\textwidth, clip, trim=2.7cm 1.8cm 2.7cm 2.8cm]
       {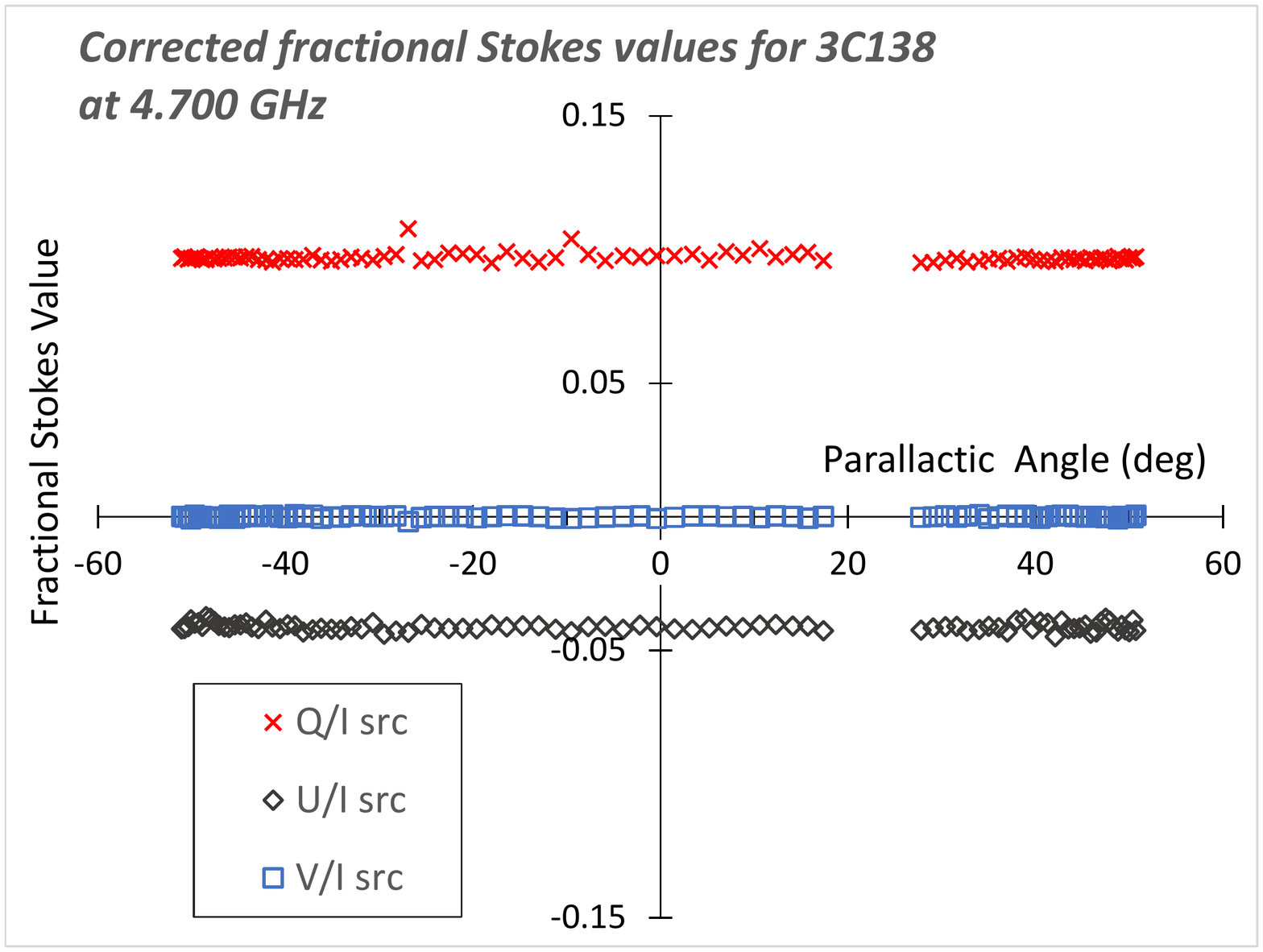}}
    \end{center}
    \caption{Source Stokes parameters for (a) 3C286 and (b) 3C138 calculated by 
    applying equation (\ref{eqn6}) to the observed values in Figures~\ref{Fit} 
    and \ref{Fit2}. The 3C286 data shows some noise at P.A.s above and below 
    $40\degr$, this noise is seen in the observed fractional Stokes parameters 
    in Figure~\ref{Fit}.
    \label{Fig:srcSTKs}}
\end{figure}

\begin{figure} [t]
    \begin{center}
    \subfigure[]{
       \includegraphics[width=0.49\textwidth, clip, trim=1.8cm 2.0cm 1.8cm 2.6cm]
       {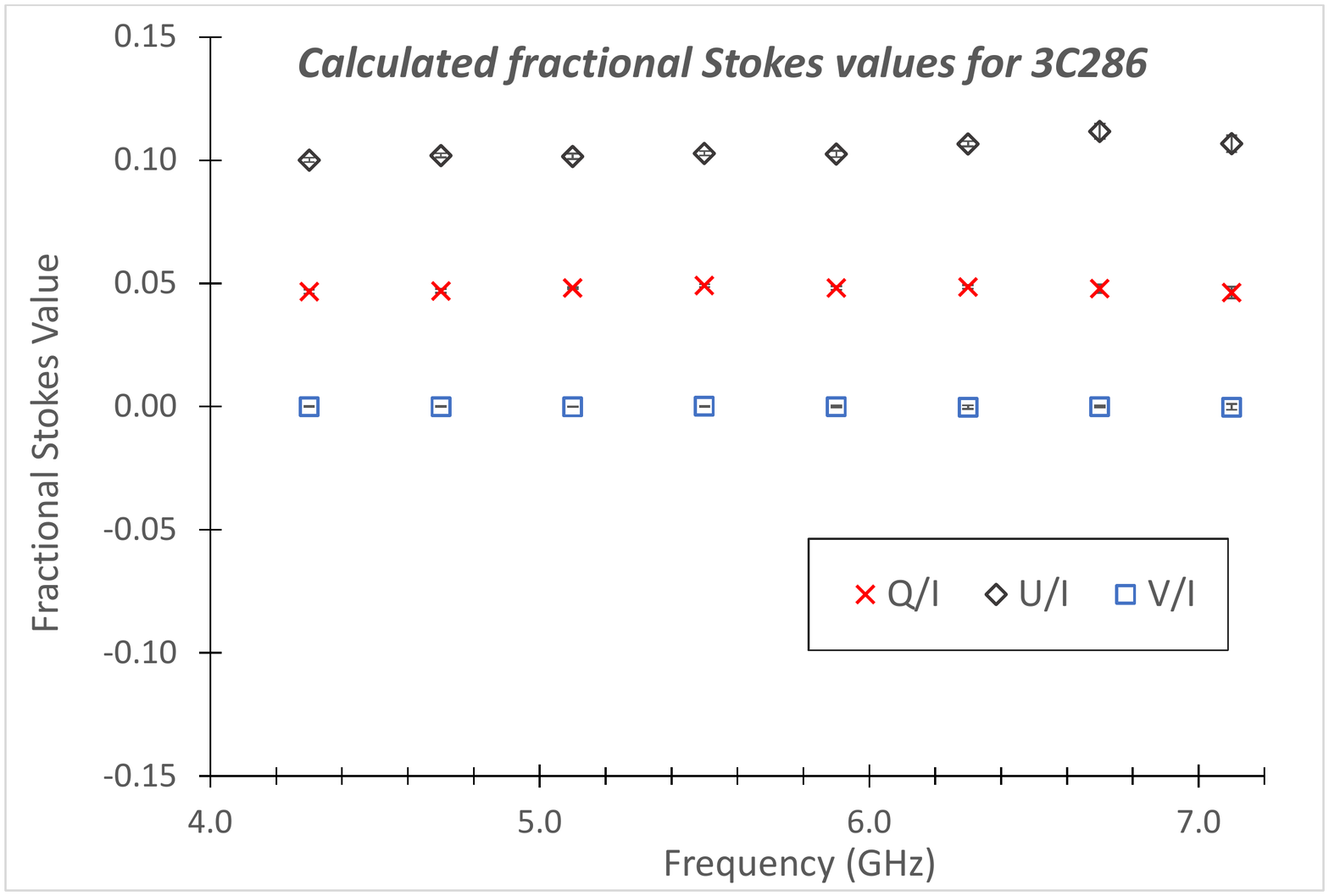}}          
    \subfigure[]{
       \includegraphics[width=0.49\textwidth, clip, trim=1.8cm 2.0cm 1.8cm 2.6cm]
       {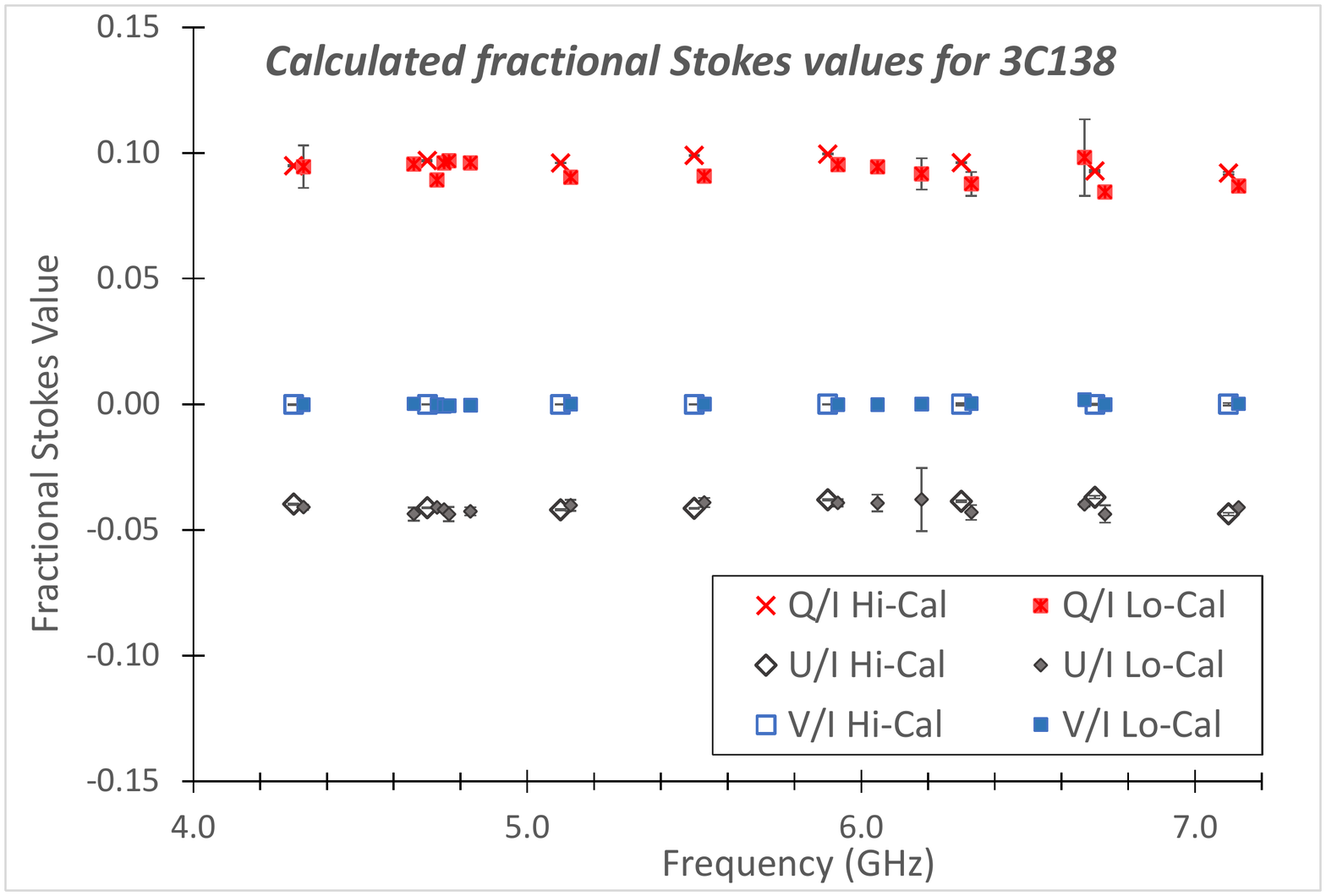}}
    \end{center}
    \caption{Source Stokes parameters for (a) 3C286 and (b) 3C138 for the observed 
    C-Band frequencies. The points are average values from the frequency set of 
    Spider scans. The error bars are the standard error of the mean and, for the 
    Spider scan (Hi-Cal ND) data, are smaller than the plotted symbols; because 
    the top and bottom of the error bar are almost on top of one another in some 
    cases, the error bar appears as a single horizontal line. Figure (b) includes 
    values from the Lo-Cal ND calibrator observations. Note that the Lo-Cal data 
    points for the evenly spaced frequencies have been offset to the right by 30 
    MHz so that they do not lie on top of the Hi-Cal points.
    \label{Fig:freqSTKs}}
\end{figure}

\rvw{Application of the derived Mueller matrices using, equation (\ref{eqn6}),
to the observed data in Figures~\ref{Fit} and \ref{Fit2} gives the source 
Stokes parameters presented in Figure~\ref{Fig:srcSTKs}. The calculated 
source values show no statistically significant variation with P.A. The 
source Stokes parameters determined in this manner for all the observed 
frequencies are presented in Figure~\ref{Fig:freqSTKs}, and show slight 
variation with frequency.}

The average linear polarization, determined from the source Stokes values over the 
eight frequencies, for 3C286 is $11.5\pm0.1\%$ at an angle of $32.5\pm0.1\degr$, and 
for 3C138 it is $10.39\pm0.09\%$ at an angle of $-11.4\pm0.2\degr$. These match the 
listed polarizations of 3C286 and 3C138 at these frequencies on the NRAO website\footnote{https://science.nrao.edu/facilities/vla/docs/manuals/obsguide/modes/pol 
(Table 7.2.7), and 
\\ https://science.nrao.edu/facilities/vla/docs/manuals/obsguide/modes/flux-density-scale-polarization-leakage-polarization-angle-tables  \label{fn:nrao}}. 
One of the beauties of this method \rvw{to determine the Mueller matrix} is that it makes
no assumptions about the source polarization parameters and thus provides an independent 
measure of their values.

\begin{figure} [t]
    \begin{center}
    \subfigure[]{
       \includegraphics[width=0.49\textwidth, clip, trim=1.8cm 2.0cm 1.8cm 2.6cm]
       {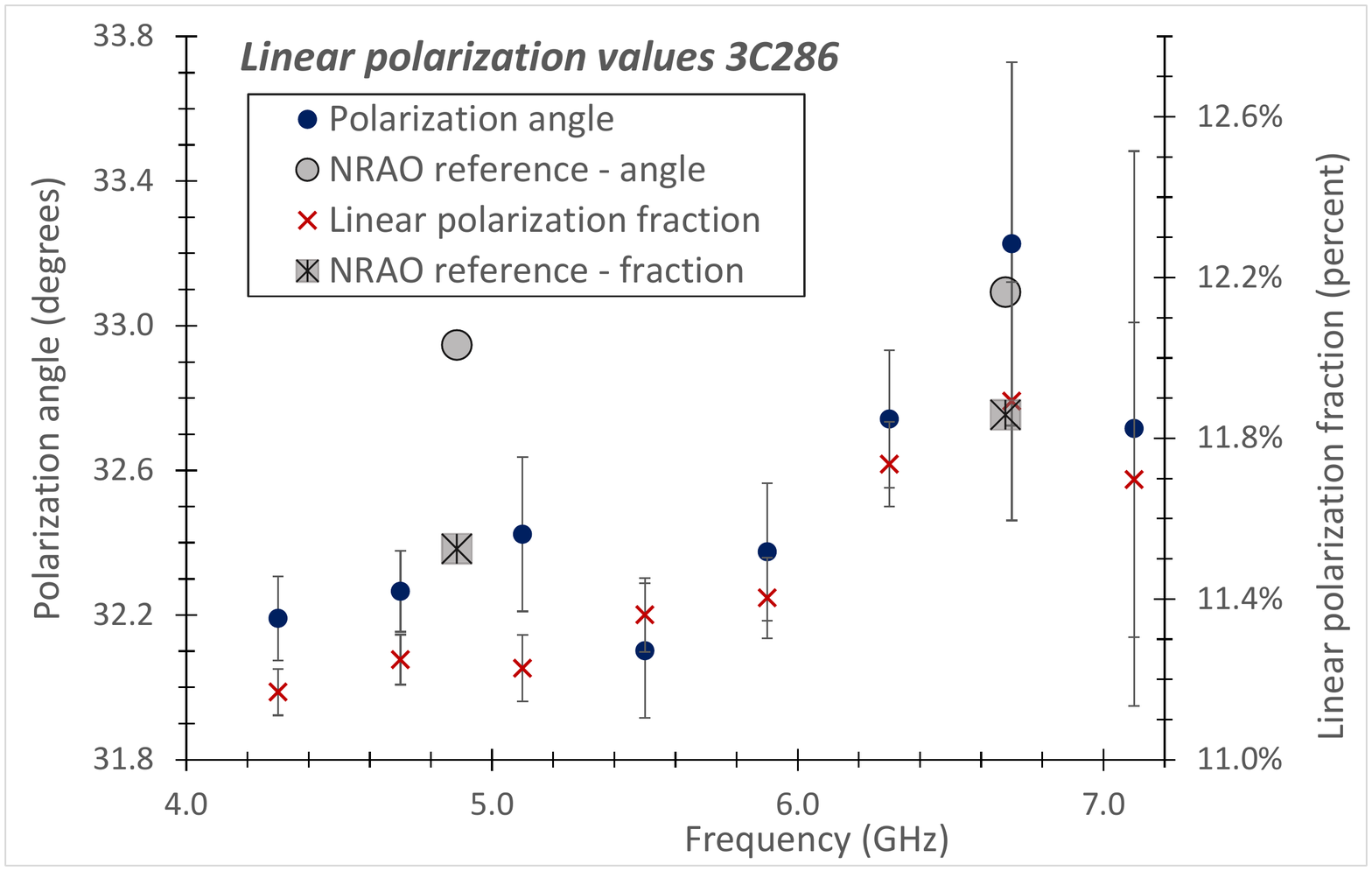}\label{Fig:FreqDepA}}          
    \subfigure[]{
       \includegraphics[width=0.49\textwidth, clip, trim=1.8cm 2.0cm 1.8cm 2.6cm]
       {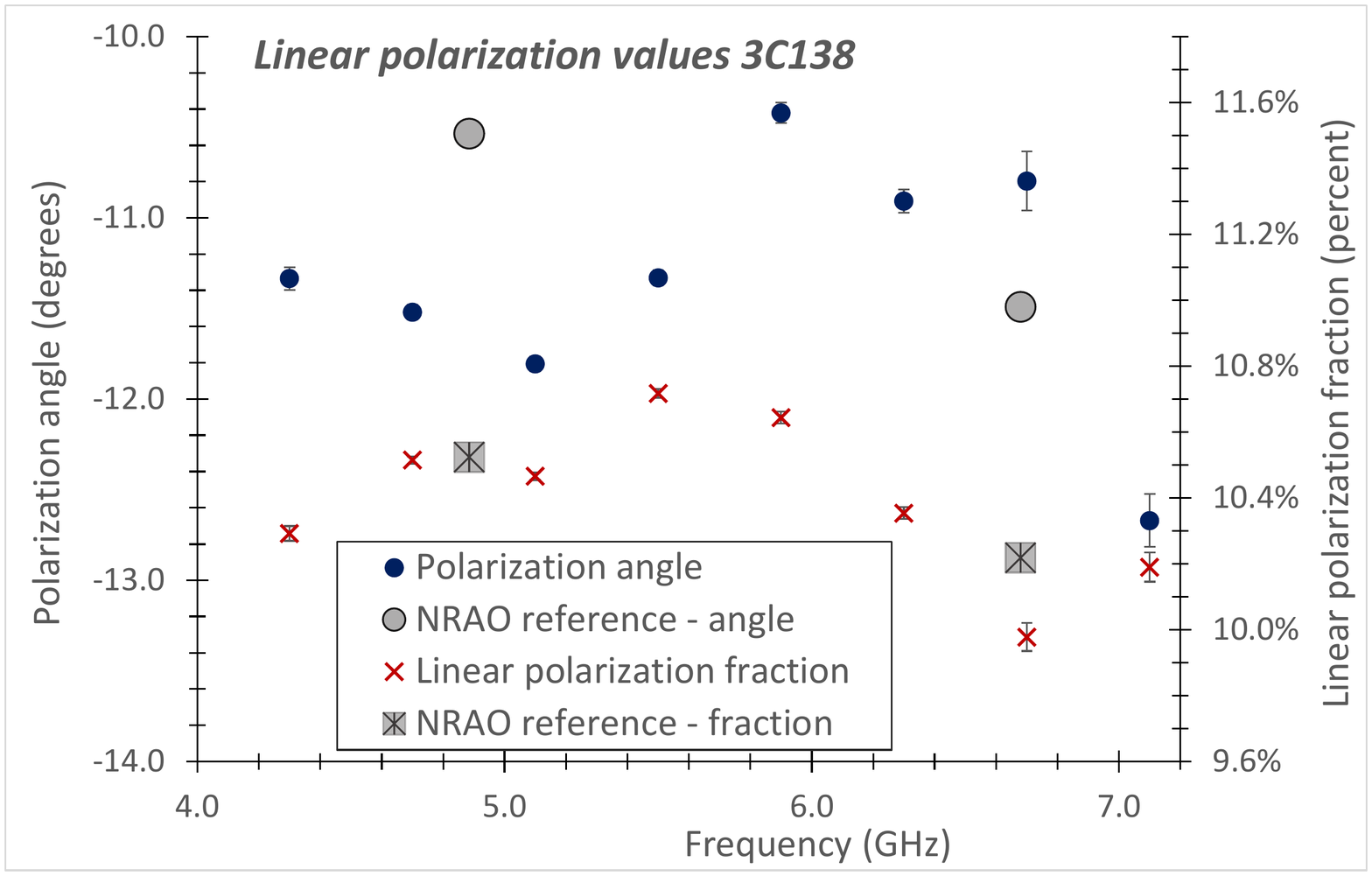}\label{Fig:FreqDepB}}
    \end{center}	
    \caption{Polarization values as a function of frequency for (a) 3C286 
    and (b) 3C138. The points are average values and the error bars are the 
    standard error of the mean from all the Spider scan observations, using 
    equation (\ref{eqn6}) to obtain source Stokes parameters. The NRAO 
    reference values are the current polarization values on the NRAO 
    website\footref{fn:nrao}. \label{Fig:FreqDep}}
\end{figure}

\rvw{The small variation of source Stokes values with frequency can be used to show 
the frequency dependence of the source polarization. The polarization angle and 
linear polarization percentage as functions of frequency for 3C286 and 3C138 are 
shown in Figure~\ref{Fig:FreqDep} (a) and (b) respectively. As can be seen from 
the plots, the polarization of 3C286 suggests a C-Band frequency dependence, 
whereas 3C138 shows no clear frequency dependent pattern across the C-Band -- 
these frequency dependencies of the polarization parameters are consistent with 
the spread of values on the NRAO website\footref{fn:nrao}.}

The Mueller matrices determined from the Spider runs are presented in 
Table~\ref{table:Mm1}. The values are the average for the separate 
observations of 3C286 and 3C138 and the uncertainty is the standard 
error of the mean. The matrices have diagonal terms close to $\pm1$ and 
relatively small off-diagonal terms.

\begin{table} [t]
\caption{GBT Spider run Mueller matrices for each of the eight observed C-Band 
frequencies. For each component of the Mueller matrix, uncertainties for the 
least significant figures are in parentheses.}
\label{table:Mm1} 
\vspace{2mm}
\begin{minipage}[c]{0.5\textwidth}
	\centering
	\textbf{4.300 GHz} \\ [1.0ex] $\left[
	\begin{tabular}{r@{.}lr@{.}lr@{.}lr@{.}l} 
          1&0000(0)  &   0&0440(25) & --0&0010(4)  & --0&0052(4) \\
        --0&0441(25) & --1&0000(0)  &   0&0000(0)  &   0&0068(23) \\
          0&0015(4)  & --0&0007(3)  & --0&9941(2)  & --0&1082(22) \\
        --0&0048(2)  &   0&0068(23) & --0&1082(22) &   0&9941(3) \\
	\end{tabular} \right]$ 
\end{minipage}
\begin{minipage}[c]{0.5\textwidth}
	\centering
	\textbf{5.900 GHz} \\ [1.0ex] $\left[
	\begin{tabular}{r@{.}lr@{.}lr@{.}lr@{.}l}
        1&0000(0)  & --0&0289(44)  & --0&0035(2)  &   0&0002(9) \\
        0&0289(44) & --0&9997(3)   &   0&0000(0)  &   0&0172(178) \\
        0&0035(3)  & --0&0013(14)  & --0&9973(2)  & --0&0729(33) \\
        0&0000(13) &   0&0171(178) & --0&0729(34) &   0&9970(6) \\
	\end{tabular} \right]$
\end{minipage}
\\
\\ [4.0ex]
\begin{minipage}[c]{0.5\textwidth}
	\centering
	\textbf{4.700 GHz} \\ [1.0ex] $\left[
	\begin{tabular}{r@{.}lr@{.}lr@{.}lr@{.}l}
          1&0000(0)  & --0&0009(35) &   0&0020(14) & --0&0007(0) \\
          0&0009(35) & --1&0000(0)  &   0&0000(0)  & --0&0004(1) \\
        --0&0019(14) &   0&0000(0)  & --0&9946(3)  & --0&1042(28) \\
        --0&0009(1)  & --0&0004(1)  & --0&1042(28) &   0&9946(3) \\
	\end{tabular} \right]$
\end{minipage}
\begin{minipage}[c]{0.5\textwidth}
	\centering
	\textbf{6.300 GHz} \\ [1.0ex] $\left[
	\begin{tabular}{r@{.}lr@{.}lr@{.}lr@{.}l} 
          1&0000(0)  & --0&0231(16)  & --0&0008(2)  & --0&0035(3) \\
          0&0231(18) & --0&9982(1)   &   0&0000(0)  & --0&0017(596) \\
          0&0012(3)  &   0&0005(59)  & --0&9950(5)  & --0&0992(51) \\
        --0&0033(11) & --0&0016(593) & --0&0994(51) &   0&9933(6) \\
	\end{tabular} \right]$
\end{minipage}
\\ 
\\ [4.0ex]
\begin{minipage}[c]{0.5\textwidth}
	\centering
	\textbf{5.100 GHz} \\ [1.0ex] $\left[
	\begin{tabular}{r@{.}lr@{.}lr@{.}lr@{.}l}
          1&0000(0)  &   0&0003(25) &   0&0056(7)  & --0&0005(2) \\
        --0&0003(25) & --1&0000(0)  &   0&0000(0)  &   0&0085(31) \\
        --0&0055(7)  & --0&0008(3)  & --0&9960(3)  & --0&0888(31) \\
        --0&0010(1)  &   0&0084(30) & --0&0888(31) &   0&9960(3) \\
	\end{tabular} \right]$
\end{minipage}
\begin{minipage}[c]{0.5\textwidth}
	\centering
	\textbf{6.700 GHz} \\ [1.0ex] $\left[
	\begin{tabular}{r@{.}lr@{.}lr@{.}lr@{.}l}
          1&0000(0)  &   0&0386(76)  & --0&0049(33)  &   0&0040(20) \\
        --0&0385(77) & --0&9991(9)   &   0&0000(0)   &   0&0240(361) \\
          0&0048(34) & --0&0014(18)  & --0&9991(5)   & --0&0406(119) \\
          0&0048(9)  &   0&0240(361) & --0&0406(119) &   0&9982(14) \\
	\end{tabular} \right]$
\end{minipage}
\\ 
\\ [4.0ex]
\begin{minipage}[c]{0.5\textwidth}
	\centering
	\textbf{5.500 GHz} \\ [1.0ex] $\left[
	\begin{tabular}{r@{.}lr@{.}lr@{.}lr@{.}l}
          1&0000(0)  &   0&0189(27) & --0&0003(8)  &   0&0000(2) \\
        --0&0189(26) & --0&9999(0)  &   0&0000(0)  &   0&0136(5) \\
          0&0003(8)  & --0&0011(0)  & --0&9966(1)  & --0&0821(14) \\
          0&0003(3)  &   0&0136(5)  & --0&0821(14) &   0&9965(1) \\
	\end{tabular} \right]$
\end{minipage}
\begin{minipage}[c]{0.5\textwidth}
	\centering
	\textbf{7.100 GHz} \\ [1.0ex] $\left[
	\begin{tabular}{r@{.}lr@{.}lr@{.}lr@{.}l} 
          1&0000(0)  & --0&0181(21)  &   0&0028(47)  & --0&0036(5) \\
          0&0182(19) & --0&9993(7)   &   0&0000(0)   & --0&0230(283) \\
        --0&0027(46) &   0&0000(2)   & --0&9996(4)   & --0&0215(170) \\
        --0&0033(1)  & --0&0230(283) & --0&0215(170) &   0&9990(3) \\
	\end{tabular} \right]$
\end{minipage}
\\ [0.1ex]
\end{table}

The results presented here only use data from when the source was at the 
center of the beam (i.e.~when the signal was at maximum), and when the beam 
was pointed off-source (to determine the background signal level). The rest 
of the Spider scan data can be used to determine the polarization parameters 
across the whole beam of the telescope. Analysis of the half-power points 
averaged over all the Spider scan data produces Mueller matrices similar to 
the beam-center results. Further analysis of the off-axis 
signal data has not been attempted here. 

\rvw{
\subsection{Frequency dependence of the Mueller Matrix Parameters}\label{sec:MMpara}}
\rvw{In Figures \ref{Fig:Delta-G} -- \ref{Fig:phi}, the five Mueller matrix 
parameters derived in determining the matrices in Tables \ref{table:Mm1}, 
\ref{table:MM_Lo-Cal1} and \ref{table:MM_Lo-Cal_maser} 
are presented as functions of frequency.}

\rvw{
The $\Delta G$ parameter is the relative calibration gain difference 
between the $X$ and $Y$ components of the feed and is the basis of the 
$m_{IQ}$ and $m_{QI}$ Mueller matrix terms (see Appendix \ref{MMeq}). 
It has a dominant impact when applying the Mueller matrix to observations 
of sources with partial linear polarization, in particular when Stokes $Q$ is
substantially smaller than Stokes $I$. The $\Delta G$ values obtained from 
our observations differ between the Hi-Cal and Lo-Cal NDs, as seen in Figure 
\ref{Fig:Delta-G}, resulting in different Mueller matrices required for the 
two NDs, as described in Subsection \ref{sec:HiLo}.}

\rvw{
In addition, the $\Delta G$ values do not show any analytically predictable 
frequency dependence across the range of the new C-Band receiver, although 
such a variation was reported for the old GBT C-Band receiver 
\citep{Heiles_et_al_2003}. Because the Mueller matrix is sensitive to the 
value of $\Delta G$, using slightly different $\Delta G$ values with differences 
less than the dispersion seen in Figure \ref{Fig:Delta-G} results in inaccurate 
polarisation values. Interpolating values of $\Delta G$ for specific transition 
frequencies using values from the matrices in Table \ref{table:MM_Lo-Cal1} 
results in polarisation angle error spread of $\pm8\degr$ and a linear 
polarisation fraction dispersion of $\pm2\%$. The polarisation results from a 
linear fit of $\Delta G$ with frequency show an even broader dispersion. 
Subsequent separate Spider observations (T. Robishaw, private communication) 
for 64 different C-Band frequencies displayed similar $\Delta G$ values and 
frequency dependence confirming the $\Delta G$ frequency behaviour as real 
and not just measurement uncertainty. 

\rvw{The $\psi$ parameter changes with frequency, as shown in Figure~\ref{Fig:psi}. 
Because $\psi$ is the phase difference between the calibration ND signal and 
the incoming radiation (Appendix \ref{MMeq}), a linear frequency dependence is 
expected due to path length differences. The $\psi$ frequency derivative, from 
the linear fit in Figure~\ref{Fig:psi}, implies a difference of about 1 mm between 
the calibration ND signal and the incoming radiation, which is reasonable for the 
components involved. The Hi-Cal and Lo-Cal ND data have slightly different slopes, 
but this difference is not statistically significant.} 
\begin{figure} [t]
	\begin{center}
    \subfigure[]{
       \includegraphics[width=0.48\textwidth, clip, trim=2.0cm 1.8cm 2.0cm 1.8cm]
       {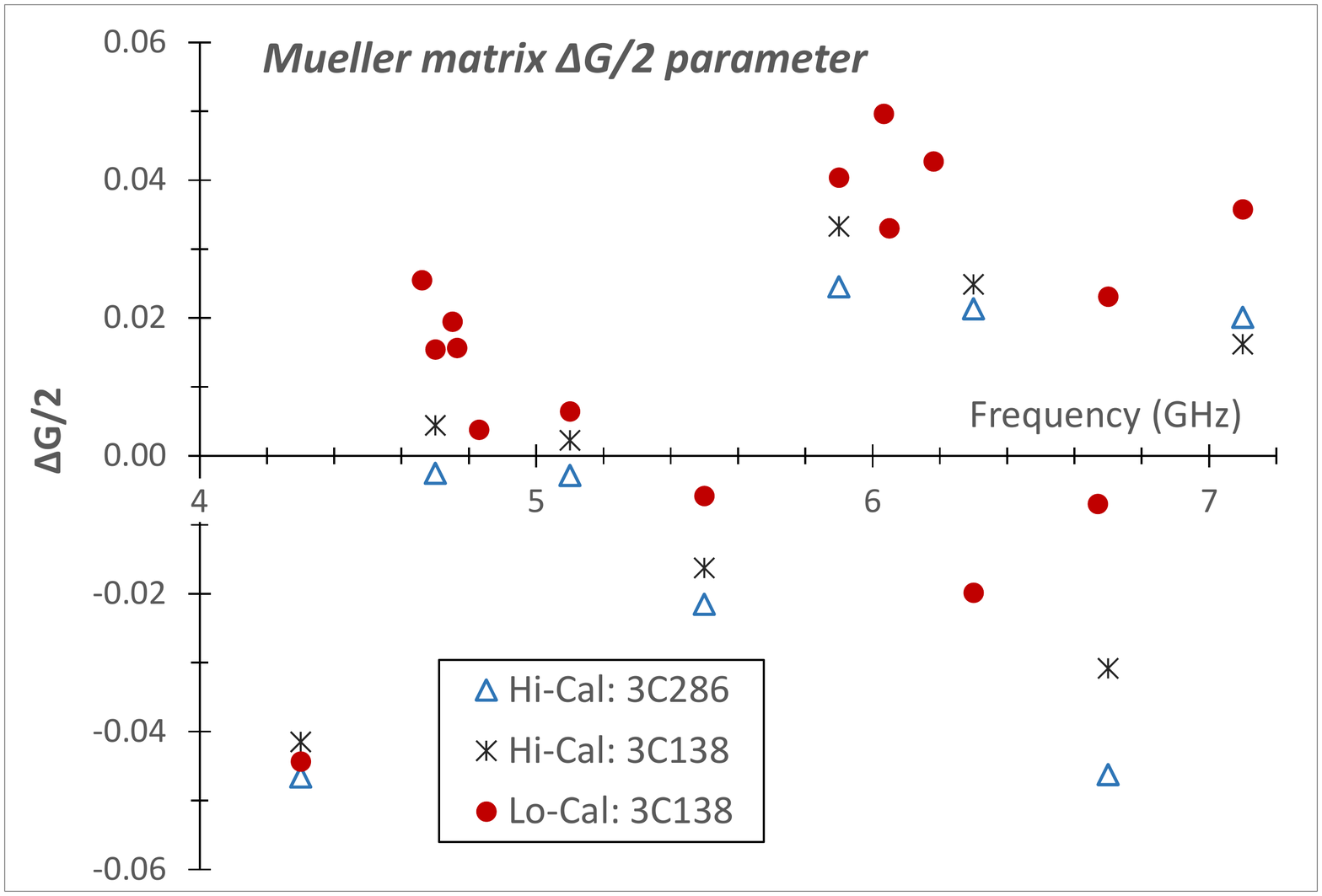} \label{Fig:Delta-G}}
    \subfigure[]{
       \includegraphics[width=0.48\textwidth, clip, trim=2.0cm 1.8cm 2.0cm 1.8cm]
       {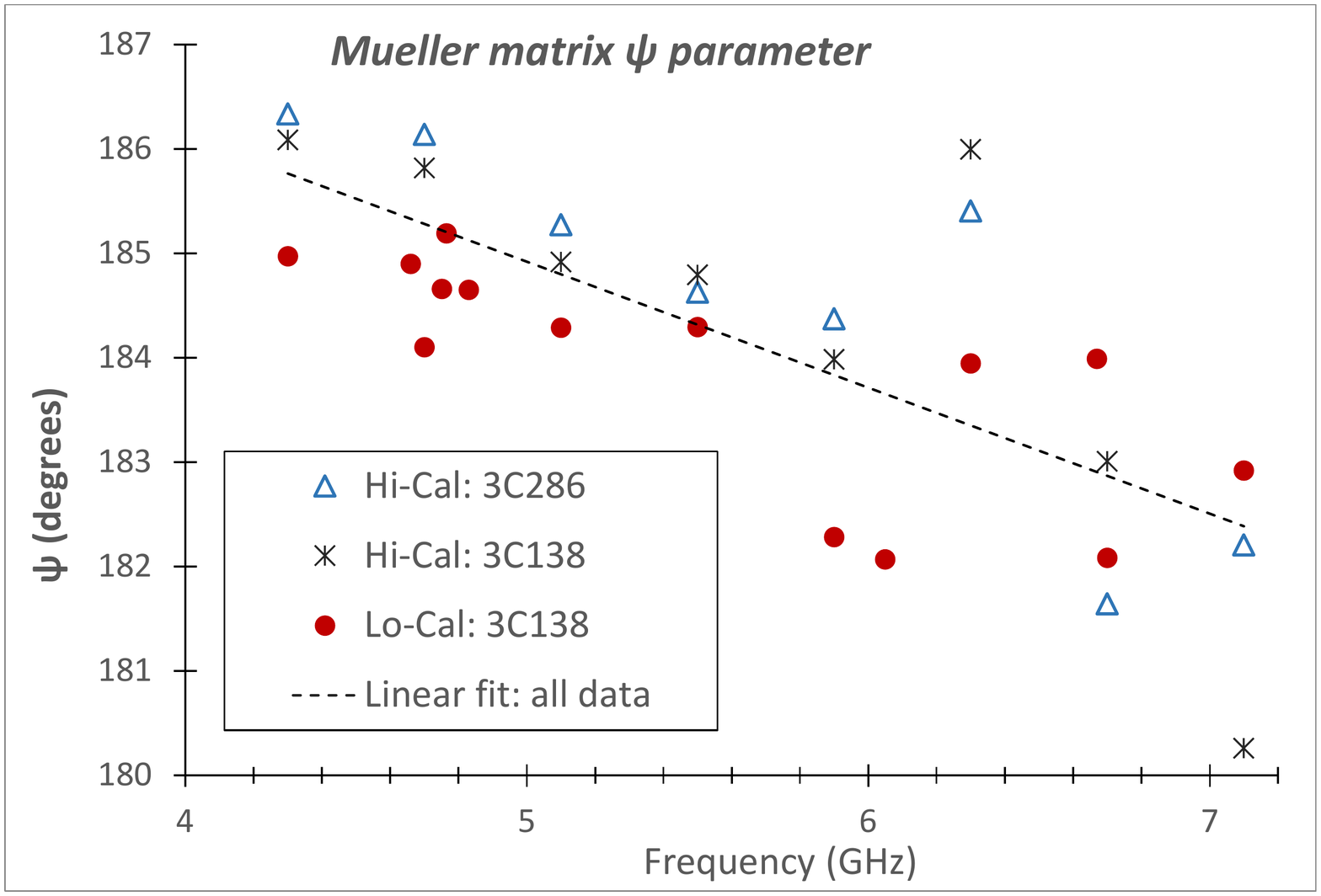} \label{Fig:psi}}
    \end{center}
    \subfigure[]{
       \includegraphics[width=0.48\textwidth, clip, trim=2.0cm 1.8cm 2.0cm 1.8cm]
       {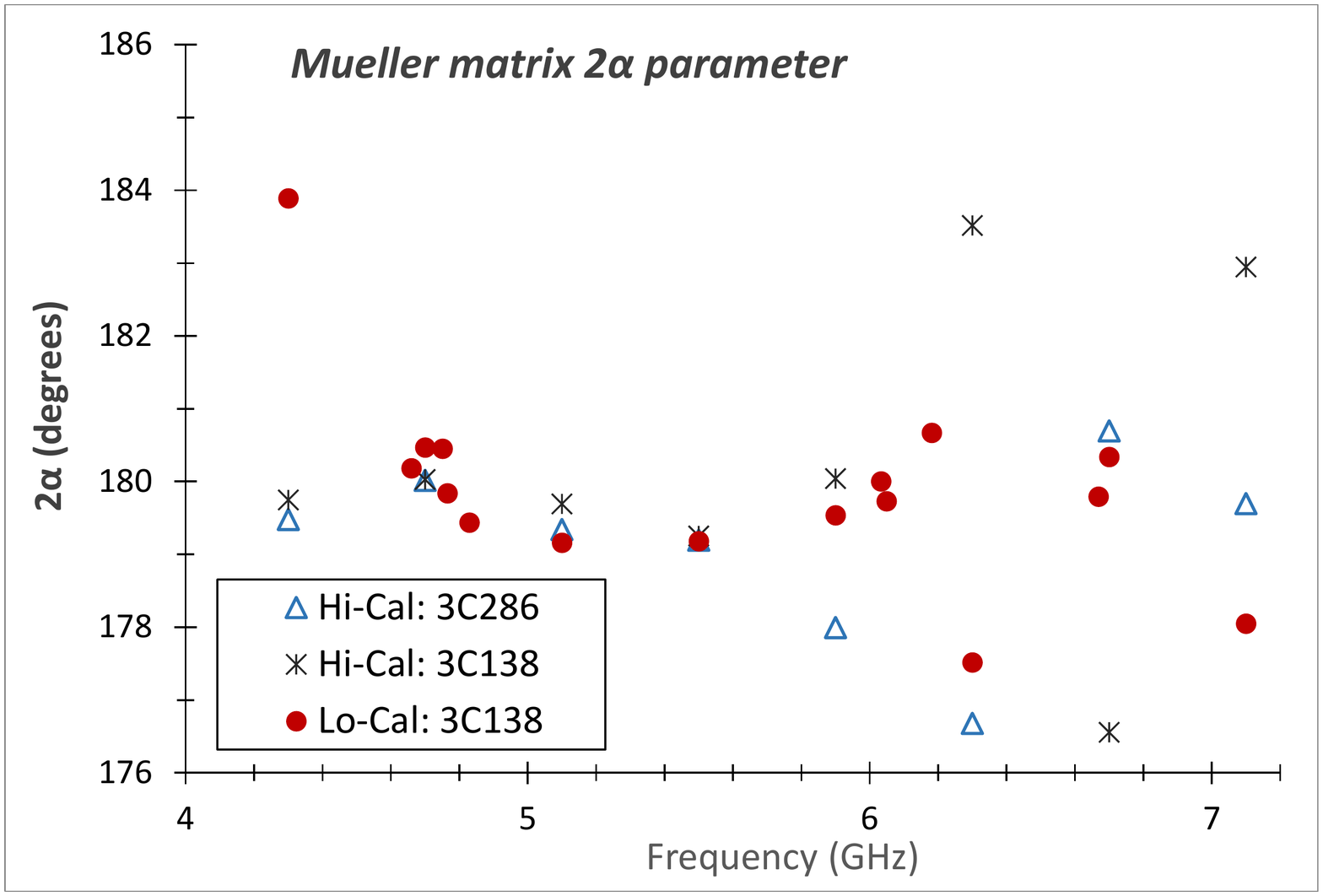} \label{Fig:alpha}}
    \subfigure[]{
       \includegraphics[width=0.48\textwidth, clip, trim=2.0cm 1.8cm 2.0cm 1.8cm]
       {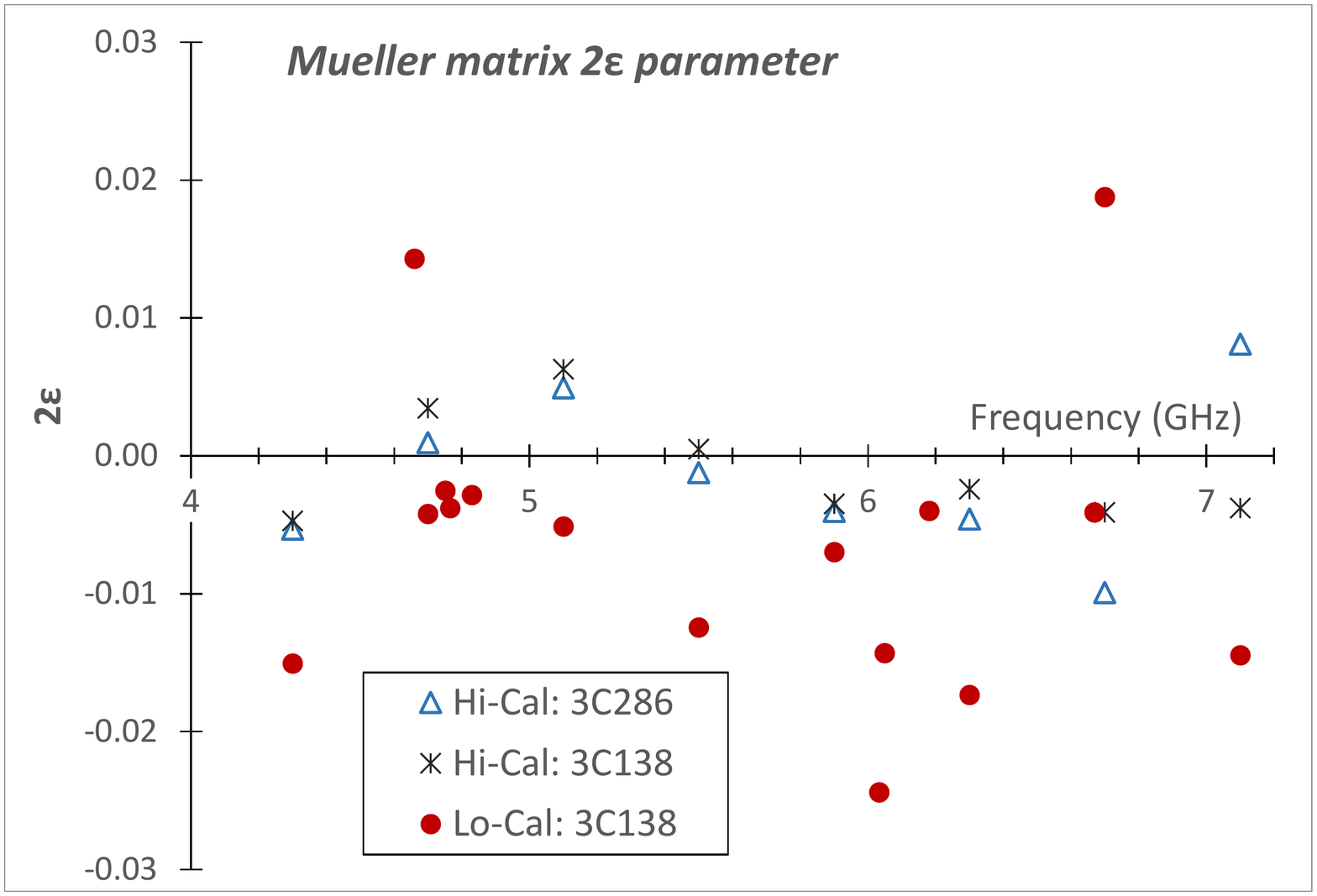} \label{Fig:epsilon}}
    \subfigure[]{
       \includegraphics[width=0.48\textwidth, clip, trim=2.0cm 1.8cm 2.0cm 1.8cm]
       {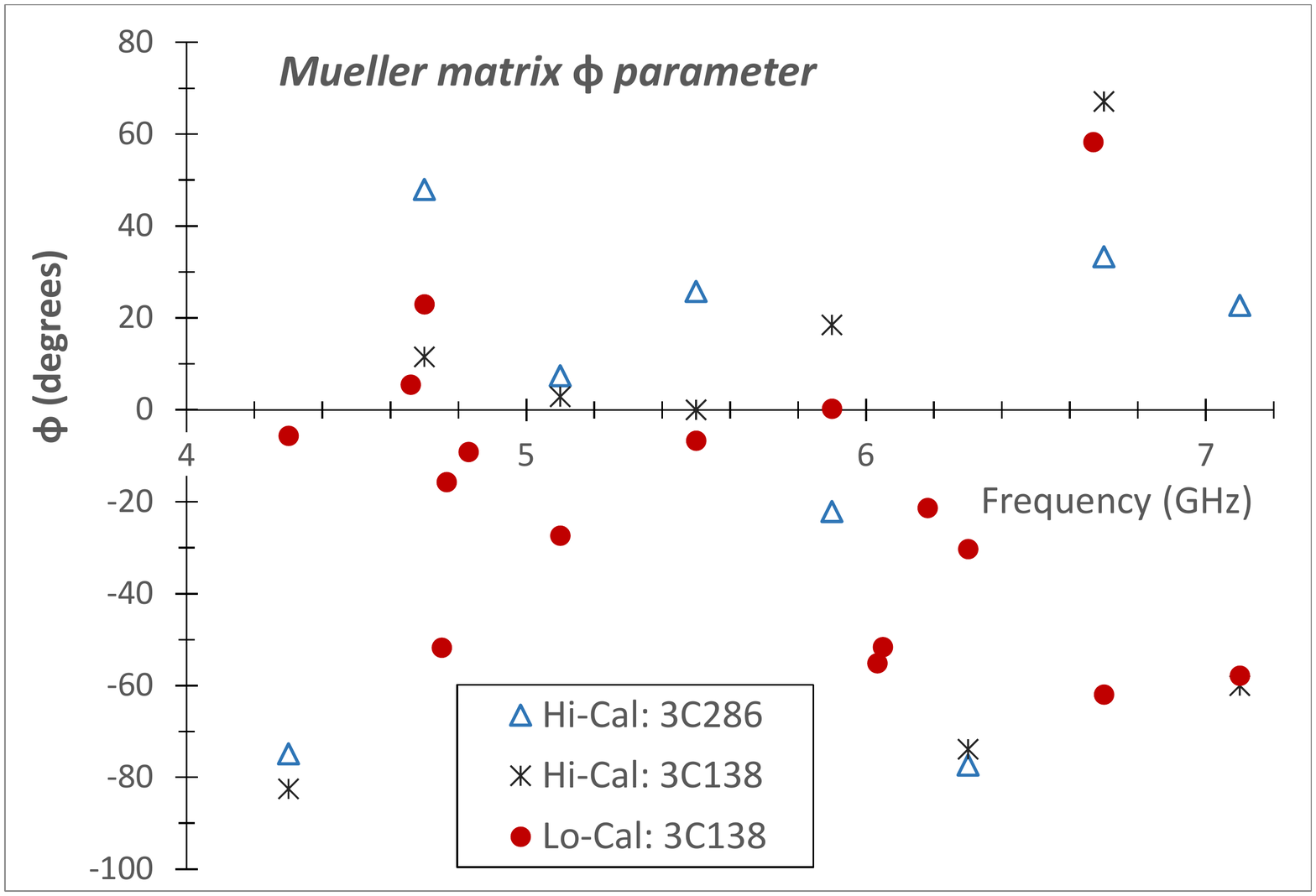} \label{Fig:phi}}
    \caption{Mueller matrix parameters as a function of frequency. 
        (a) $\Delta G/2$, (b) $\psi$, with a linear best fit through all the 
        data points. The linear fit has values of $190.95 -1.2067 
        \times \text{Frequency (GHz)}$,
        (c) $2\alpha$, (d) $2\epsilon$, and (e) $\phi$.}
\end{figure}

\rvw{From Figure \ref{Fig:alpha}, it can be seen that the $2\alpha$ parameter 
scatters around $180\degr$, with increasing dispersion at the high frequency 
end of the C-Band. However, it does not appear to have a frequency dependence. 
\cite{Heiles_et_al_2003} presents values for $\alpha$ over a range of 
frequencies which show no change across the C-Band, but differences between L-, 
C- and X-Band. The $\epsilon$ and $\phi$ parameters, shown in Figures 
\ref{Fig:epsilon} and \ref{Fig:phi}, have a small impact on the C-Band Mueller 
matrices. They scatter around values of 0 and $0\degr$ respectively. Any 
inaccuracy in $\Delta G$ or $\psi$ will result in additional noise or scatter 
in the fitted outcome for $\alpha$, $\epsilon$ and $\phi$.}

\rvw{
\subsection{Hi-Cal and Lo-Cal ND Mueller Matrix Differences}\label{sec:HiLo}}
\rvw{The Mueller matrix determined from our observations, differs depending 
on whether the Hi-Cal or Lo-Cal ND is used to calibrate the data.} Initial 
analysis applying the Hi-Cal Mueller matrix to observations of 3C138 made 
using the Lo-Cal ND on 2021 January 05 produced inaccurate polarization 
outcomes. Another set of observations made using the Lo-Cal ND in 2021 
November produced results similar to the January set, creating the 
realisation that different Mueller matrices are required for observations with 
the Lo-Cal or Hi-Cal ND. This led to derivation of the Lo-Cal ND Mueller matrix 
from observations of 3C138 at different P.A.s. Because the fractional source 
Stokes parameters for 3C138 are known from the fitting process using the 
Spider run, these values can be substituted into equations (\ref{eqnQIsrc}) 
and (\ref{eqnUIsrc}) so that equations (\ref{eqn9}) -- (\ref{eqn12}) contain 
only the five unknown parameters of the Mueller matrix. Observations of the 
polarization calibrator made before and after our programme source have 
different P.A.s and can be fitted to these equations to derive the Mueller 
matrix parameters. In Figure \ref{Fit2} the Lo-Cal ND observations, shown with 
larger symbols and dot-dashed lines, are at P.A.s of $23.5\degr$ and $47.2\degr$. 
Fractional Stokes $Q/I$ \revfour{and $U/I$} show a difference between Hi-Cal and 
Lo-Cal ND observations, highlighting the need for different Mueller matrices 
depending on which ND is used.}

Mueller matrices determined for Lo-Cal ND observations using the above method 
are presented in Table \ref{table:MM_Lo-Cal1} in Appendix B  for evenly spaced 
frequencies and in Table \ref{table:MM_Lo-Cal_maser} for the maser frequencies 
used in our programme observations. 
\revfour{There are differences between the Hi-Cal and Lo-Cal ND Mueller matrices
parameters but the differences are not significant at all the observed frequencies.}  
The differences, an example of which is shown in Table \ref{table:MM-Lo_vs_Hi} 
for observations at 4.700 GHz, are mainly due to the $\Delta G$ parameter.

\clearpage
\revfour{
\subsection{Calibration of the GBT polarisation products}\label{sec:GBTcal}}
\revfour{The GBT provides calibration values for the four polarisation products 
$XX$, $YY$, $XY$ and $YX$, with seperate $T_\text{cal}$ values stored and included 
in the spectrum SDFITS header (Appendix \ref{SpectraCal}). The standard GBT 
calibration provides the same $T_\text{cal}$ value for $XX$, $XY$ and $YX$ and a 
different value for $YY$. These values are calibrated from on sky observations and 
vary with frequency. However, as noted by \cite{Goddy_et_al_2020}, these values 
have not been regulary updated and are only expected to be accurate at the $15 - 20\%$ 
level.}

\revfour{Using the different $XX$ and $YY$ $T_\text{cal}$ calibrations 
provided in the spectrum SDFITS header resulted in larger Mueller matrix 
$\Delta G$ values than if the same $T_\text{cal}$ value was used 
for all the polarization products. In addition, the $\Delta G$ difference 
between the Hi-Cal and Lo-Cal ND Mueller matrices is exacerbated. As a result, 
all the analysis in this paper has been done using the same $T_\text{cal}$ 
value for all four polarization products, and thus use of the Mueller 
matrices presented in Tables \ref{table:Mm1},\ref{table:MM-Lo_vs_Hi}, 
\ref{table:MM_Lo-Cal1} and \ref{table:MM_Lo-Cal_maser} only apply to spectra 
calibrated in the same manner.}

\revfour{In the next section, a method is described to modify $\Delta G$ to 
adjust for inaccuracy in the relative $XX$ and $YY$ $T_\text{cal}$ calibration, 
so that a generic, frequency-independent Mueller matrix can be utilized. }

\revfour{
\section{Frequency independent Mueller matrix}
\label{sec:FreqInd}}
\revfour{If $\Delta G$ can be determined independently then the spectra can be
corrected for the relative calibration gain prior to determination and use of 
the Mueller matrix. This correction removes the unpredictable $\Delta G$ change 
across the C-Band and does away with the need for bespoke ND and frequency 
Mueller matrices.}

\revfour{
\subsection{Independent calculation of the relative calibration gain}
\label{sec:Cal}}
\revfour{The relative calibration gain is defined as 
\begin{eqnarray}
XX_{\text{calibrated,modified}} & = & XX_{\text{calibrated}} 
(1 - \Delta G /2) \label{eqn:XXmod}\\
YY_{\text{calibrated,modified}} & = & YY_{\text{calibrated}} 
(1 + \Delta G /2) \label{eqn:YYmod}.
\end{eqnarray}
where $XX_{\text{calibrated}}$ and $YY_{\text{calibrated}}$ are the 
calibrated observed spectra while $XX_{\text{calibrated,modified}}$ and 
$YY_{\text{calibrated,modified}}$ are the expected polarization spectra 
should no instrument correction be required \citep{Heiles_et_al_2001}. 
These modified spectra, when converted using equations (\ref{eqn1}) and 
(\ref{eqn2}) into Stokes parameters, describe the sky rotated source signal 
$I_\text{src}$ and $Q_\text{src,rot}$.}

\revfour{
From equations (\ref{eqn:XXmod}) and (\ref{eqn:YYmod}) the relative 
calibration gain can be expressed as 
\begin{align}
\Delta G / 2 = [(Q/I)_\text{obs} - (Q/I)_\text{src,rot}] / 
[1 - (Q/I)_\text{src,rot} (Q/I)_\text{obs}] \label{eqn:Delta_G}.
\end{align}
For a source with known polarization and thus known $(Q/I)$ and $(U/I)$, 
the $(Q/I)_\text{src,rot}$ can be determined from equation (\ref{eqnQIsrc}). 
Note that the fractional Stokes values can be calculated from the 
polarization angle and the linear polarization percentage which are 
available for all the well-known polarization calibrators. This allows 
$\Delta G$ to be determined from observations of a source with known 
polarization.}

\revfour{The $\Delta G$ values were determined for the Spider scan 
observations of 3C286, using the source Stokes parameters presented 
in Section \ref{sec:results} and applying equation (\ref{eqn:Delta_G}). 
These $\Delta G$ values show no statistically significant variation 
across the P.A. observations and the average value (per frequency) is 
the same as the frequency specific Mueller matrix derived values shown 
in Figure \ref{Fig:Delta-G}. As discussed in Sections \ref{sec:MMpara} 
and \ref{sec:HiLo}, the $\Delta G$ values do vary with frequency and 
with use of either the Hi-Cal or Lo-Cal ND. }

\revfour{
\subsection{Generic GBT Mueller matrix}\label{sec:GenericMM}}
\revfour{Once the $\Delta G$ value is determined for a specific set of 
observing conditions, the observed $XX$ and $YY$ auto-correlation 
polarization products can be modified using equations (\ref{eqn:XXmod}) and 
(\ref{eqn:YYmod}). Alternatively, rewriting equation (\ref{eqn:Delta_G}) gives  
\begin{align}
(Q/I)_\text{modified} = (Q/I)_\text{src,rot} =
\frac{1}{1-(\Delta G / 2)(Q/I)_\text{obs}}
[(Q/I)_\text{obs} - \Delta G / 2] \label{eqn:Q/Imod},
\end{align}
allowing the observed fractional Stoke $(Q/I)$ value to be modified. Mueller 
matrices fitted to the modified 3C286 data result in a zero $\Delta G$ 
parameter with the other Mueller matrix parameters unchanged from the values 
shown in Figures \ref{Fig:psi}, \ref{Fig:alpha}, \ref{Fig:epsilon} and 
\ref{Fig:phi}. These Mueller matrices are independent of the frequency 
and ND variability of the $\Delta G$ parameter.}

\revfour{The Mueller matrix applicable to the $\Delta G$ modified data, can be 
described by a zero $\Delta G$ and the $\psi$ parameter determined from the 
linear function shown in Figure \ref{Fig:psi}. The $2\alpha$ parameter can 
be set to $180\degr$, and the $\epsilon$ and $\phi$ parameters to zero, which 
are reasonable values for these parameters, as seen in Figures \ref{Fig:alpha}, 
\ref{Fig:epsilon} and \ref{Fig:phi}. The scatter in these parameters has 
negligible impact on the calculated polarization values. This generic Mueller 
matrix has the form
\begin{align}
 \boldsymbol{M_\textrm{Generic Mueller}} = 
\begin{bmatrix}
    1 &  0 &          0 &           0 \\
    0 & -1 &          0 &           0 \\
    0 &  0 & \cos{\psi} &  \sin{\psi} \\
    0 &  0 & \sin{\psi} & -\cos{\psi}
\end{bmatrix},
\end{align}
where $\psi = 190.95 -1.2067 \times \text{Frequency~(GHz)}$. Provided that spectra 
have been modified for $\Delta G$ as described in Section \ref{sec:Cal}, this 
generic Mueller matrix can be applied to determine the polarization of sources 
whose spectra have been observed with the GBT.}

\begin{figure} [t]
	  \begin{center}
        \subfigure[]{
           \includegraphics[width=0.49\textwidth, clip, trim=1.8cm 2.0cm 1.8cm 2.6cm]
           {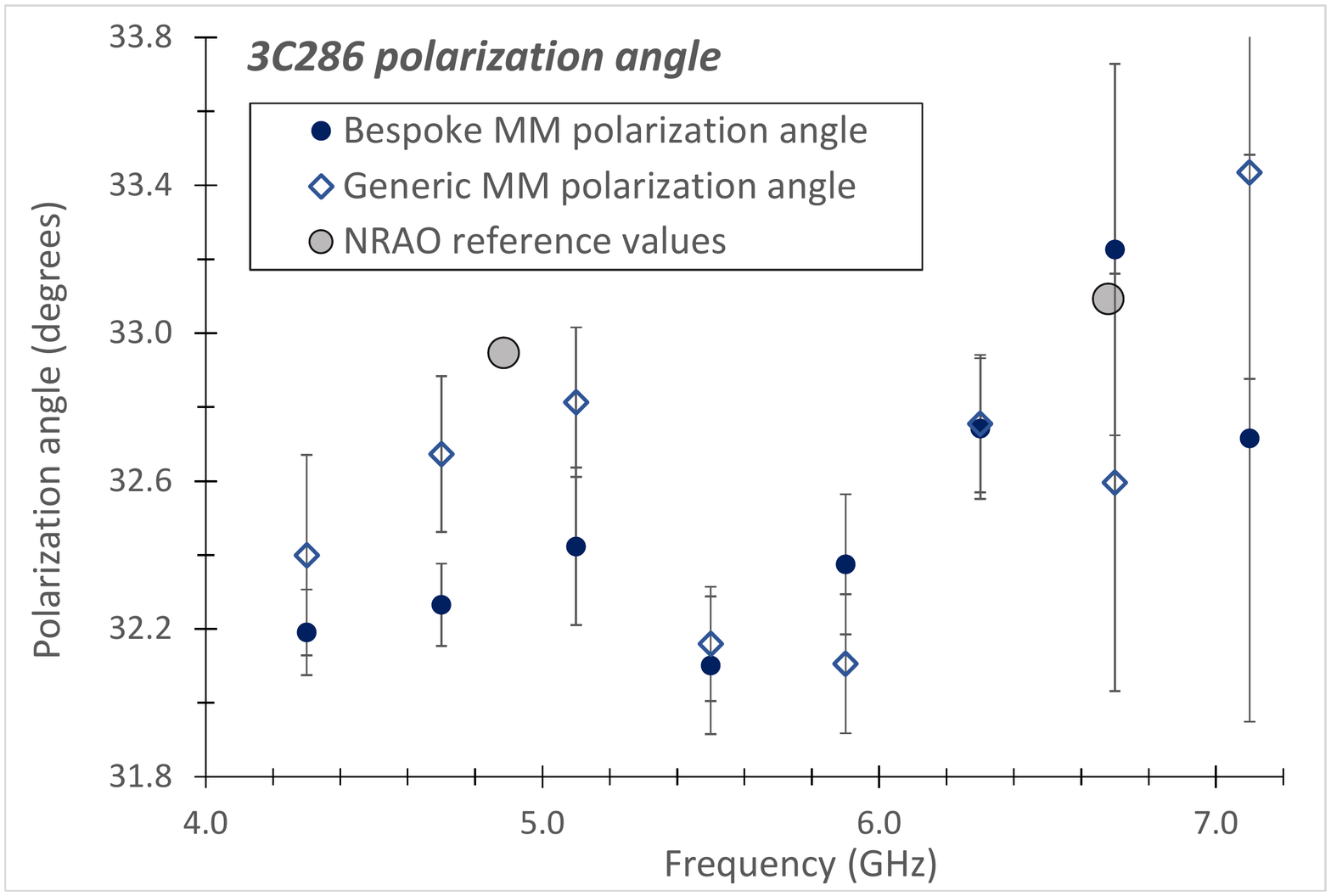}}          
        \subfigure[]{
           \includegraphics[width=0.49\textwidth, clip, trim=1.8cm 2.0cm 1.8cm 2.6cm]
           {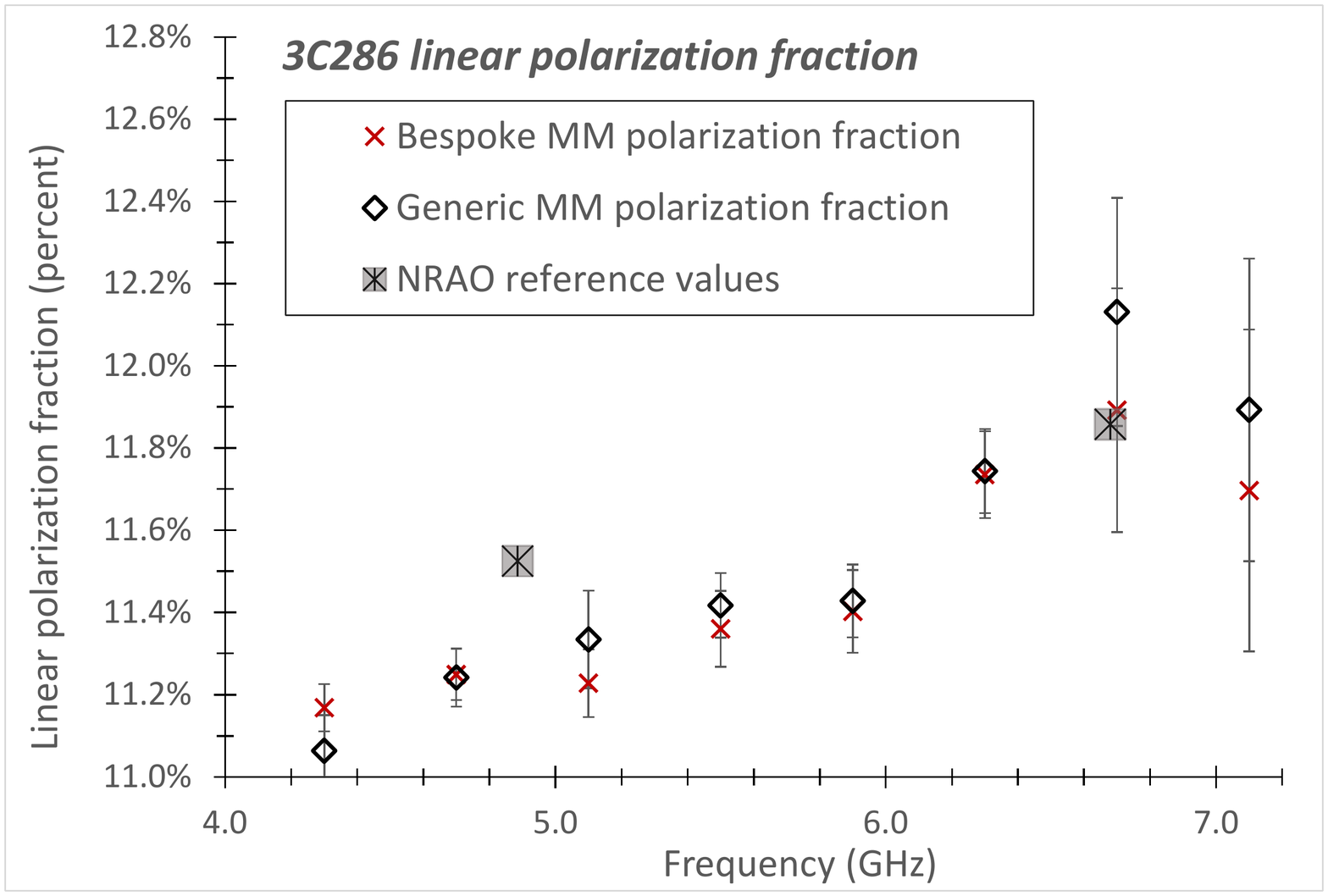}}
        \end{center}
        \caption{Polarization values for 3C286 as a function of frequency 
        (a) polarization angle and (b) linear polarization fraction. The 
        graphs compare the frequency bespoke Mueller matrix analysis 
        outcomes shown in Figure \ref{Fig:FreqDepA} with the results of 
        using the generic Mueller matrix to determine the polarization. 
        \label{Fig:3C286pol}}
\end{figure}

\revfour{
\subsection{Testing the generic GBT Mueller matrix}\label{sec:TestGenericMM}}
\revfour{This generic Mueller matrix has been applied to the $\Delta G$ 
modified 3C286 Spider scan data and the resulting polarization values, shown 
in Figure \ref{Fig:3C286pol}, are closely aligned and generally within the 
error range of the frequency bespoke Mueller matrix polarization results.} 

\revfour{The 3C138 Spider scan data has also been tested by determining the 
$\Delta G$ correction, using this to modify the data and applying the generic 
Mueller matrix to determine the polarization properties. The resulting 
polarisation values are within the range determined by use of the bespoke 
Mueller matrices described in Section \ref{sec:results} and shown in Figure 
\ref{Fig:FreqDepB}. The Lo-Cal ND observation, described in Section 
\ref{sec:HiLo}, were tested and the polarization results are within the 
same range of outcomes calculated using the bespoke Mueller matrices.}

\revfour{This testing shows that the generic Mueller matrix produces reliable 
polarization results. It is recommended that observations to determine 
$\Delta G$ are done at two or more P.A.s, as the results improve with the 
number of different pointings, as evidenced with analysis of the Lo-Cal ND 
data.}

\section{Conclusion}
In order to measure point source polarization, C-Band Mueller matrices 
have been determined for the GBT. The Hi-Cal ND Mueller matrices are 
calculated from measurements over a range of P.A.s and, hence, have 
lower uncertainties than the Lo-Cal ND measurements which are determined 
from only two or three pointings. 

\revfour{The Mueller matrices determined using the same calibration for all
four polarization products, differ with use of either the Hi-Cal or Lo-Cal 
ND calibrator and vary with frequency. This is a result of the relative 
calibration gain or $\Delta G$ parameter not changing in a predictable 
manner with frequency, suggesting a different Mueller matrix calibration 
is required for each specific observational configuration.}

\revfour{However, observations of a source with known polarization parameters
allows for an independent determination of the relative calibration gain. 
This $\Delta G$ value can be used to correct the data prior to analysis. 
The modified data enables a generic frequency and ND independent Mueller matrix 
to be used in the calculation of C-Band polarization. These polarization 
outcomes are shown to have accuracy similar to the polarization values 
determined using frequency and ND bespoke Mueller matrices applied to 
unmodified data.} Observations and analysis over an 18 month period indicate 
Mueller matrix stability on this timescale at the GBT.

\revfour{The analysis described in this paper could be used to determine 
Mueller matrices at other frequency bands and at other single dish telescopes.}

\begin{acknowledgments}
The authors were provided additional GBT observation sessions for project 
\textit{AGBT20B\_424} in order to determine the Mueller matrices. The 
support and guidance of the GBT scheduling staff, and the provision of 
additional observation time to do these calibrations, is greatly appreciated. 
The Green Bank Observatory is a facility of the National Science Foundation 
operated under cooperative agreement by Associated Universities, Inc.

\rvw{The guidance and support provided by Tim Robishaw, in setting up the 
Spider scan observations and providing details on determining the Mueller 
matrix, is acknowledged with gratitude. His input has been crucial for the 
observations and analysis presented in this paper.}

\rvw{We thank the anonymous referee for the thoughtful, thorough and 
constructive feedback that has provided valuable insights and improved 
the quality and comprehensiveness of the paper.}
\end{acknowledgments}

\vspace{5mm}
\facilities{GBT}

\software{GBTIDL, Excel Add-in Solver}

\newpage         
\appendix

\section{Calibration of spectra} \label{SpectraCal}
The standard process to calibrate spectra for the auto-correlation terms 
($XX$ and $YY$) using the calibration ND reference signal is
\begin{align}
\mbox{Spectrum}_{\text{data}} &= \mbox{Spectrum}_{\text{on-source}} - \mbox{Spectrum}_{\text{background}} \, (\mbox{or baseline}),
\label{eqnA1}
\\[5pt]
\mbox{Spectrum}_{\text{cal}} &= \text{Spectrum(ND}_{\text{ON}}) - \text{Spectrum(ND}_{\text{OFF}}),  
\\[5pt]
\mbox{Spectrum}_{\text{calibrated}} &= \mbox{Spectrum}_{\text{data}} \, / \, \mbox{Spectrum}_{\text{cal}} \times \mbox{calibration-constant}.
\end{align}

The calibration-constant, \revfour{to calibrate the spectrum in Kelvin 
or Janskys,} is dependent on the ND calibration temperature. 
\revfour{For the GBT, $T_\text{cal}$ values are predetermined from 
observations and stored in the header of the SDFITS file.  The standard 
GBT calibration provides the same value for $XX$, $XY$ and $YX$ and a 
different calibration value for $YY$. As described in Section 
\ref{sec:GBTcal}, all the analysis and results in this paper are based 
on using the $YY$ $T_\text{cal}$ value for all polarizations, as using 
the values provided in the SDFITS header for $XX$ and $YY$ result in 
larger Mueller matrix $\Delta G$ values. The new analysis code 
\citep{Fallon2022} can cater for either the different predetermined 
$T_\text{cal}$ values per polarization (as done in the standard GBTIDL 
routines) or applying the same $T_\text{cal}$ for all polarizations.}

The cross-polarisation $XY$ and $YX$ spectra require polar coordinates to 
derive the relative dispersive phase that the calibration signal is accumulating. 
This calibration removes the linear phase gradient resulting from the ND deflection 
from each individual pair of the original $XY$ and $YX$ spectra. 
\revfour{This calibration procedure has been implemented previously \citep{HR2022}\footnote{https://astro.berkeley.edu/$\sim$heiles/}, but is not included 
in the standard routines available at the GBT.}

The method used is a polar coordinate calibration per channel, given by\footnote{The 
$XY_{\text{data}}$ and $YX_{\text{data}}$ spectra have have their appropriate background 
or baseline removed as per equation (\ref{eqnA1}).}\textsuperscript{,}\footnote{Note 
use of the $\arctan(y,x)$ or $\text{arctan2}(y,x)$ function is crucial as this returns 
a value in the range $-180^\circ$ and $180^\circ$, versus the general 
$\arctan(single\text{-}value)$ function which only returns a value in the range 
$-90^\circ$ and $90^\circ$.}
\begin{align}
r_{\text{data}}     &= \sqrt{{(XY_{\text{data}})}^2 + {(YX_{\text{data}})}^2},
\\[5pt]
\theta_{\text{data}} &= \arctan(YX_{\text{data}},XY_{\text{data}}),
\\[5pt] 
r_{\text{cal}}      &= \sqrt{{(XY_{\text{cal}})}^2 + {(YX_{\text{cal}})}^2}, 	
\\[5pt]
\theta_{\text{cal}}  &= \arctan(YX_{\text{cal}},XY_{\text{cal}}),
\end{align}
where 
\begin{align*}
XY_{\text{cal}} &= XY\text{(ND}_{\text{ON}}) - XY\text{(ND}_{\text{OFF}}),   
\\[5pt]
YX_{\text{cal}} &= YX\text{(ND}_{\text{ON}}) - YX\text{(ND}_{\text{OFF}}),   
\end{align*}
\
and the calibrated outcome is
\begin{align}
r_\text{{calibrated}}          &= r_{\text{data}} \, / \, r_{\text{cal}},
\\[5pt]
\theta_{\text{calibrated}}      &= \theta_{\text{data}} - \theta_{\text{cal}},
\\[5pt]
XY_{\text{calibrated}} &= r_{\text{calibrated}} \times \cos (\theta_{\text{calibrated}}) 
    \times \mbox{calibration-constant},
\\[5pt]
YX_{\textrm{calibrated}} &= r_{\textrm{calibrated}} \times \sin (\theta_{\textrm{calibrated}}) 
    \times \mbox{calibration-constant}.
\end{align}

\newpage         
\section{Mueller matrix equations} \label{MMeq}
The sky rotation matrix is given by 
\begin{align}
\boldsymbol{M_\textrm{sky}} = 
\begin{bmatrix}
	1 & 0 & 0 & 0\\ 
	0 &  \cos(2 \, \mathrm{P.A.}) & \sin(2 \, \mathrm{P.A.}) & 0\\
	0 & -\sin(2 \, \mathrm{P.A.}) & \cos(2 \, \mathrm{P.A.}) & 0\\
	0 & 0 & 0 & 1
\end{bmatrix}.
\end{align}

Dividing equation (\ref{eqn5}) by $I_\textrm{src}$ gives the source fractional 
Stokes parameters, and the rotated fractional source Stokes parameters can be 
written as
\begin{eqnarray}
(Q/I)_\textrm{src,rot} & = & \cos(2 \, \mathrm{P.A.}) \, (Q/I)_\textrm{src} + 
\sin(2 \, \mathrm{P.A.}) \, (U/I)_\textrm{src} \label{eqnQIsrc} \\[5pt]
(U/I)_\textrm{src,rot} & = -&\sin(2 \, \mathrm{P.A.}) \, (Q/I)_\textrm{src} + 
\cos(2 \, \mathrm{P.A.}) \, (U/I)_\textrm{src} \ . \label{eqnUIsrc}
\end{eqnarray}

The Mueller matrix described by \citet{Heiles_et_al_2001} and \citet{Heiles2002} is
\begin{align}
\begin{bmatrix}
    1 & [-2 \epsilon \sin{\phi} \sin{2 \alpha} + (\Delta G/2) \cos{2 \alpha}] 
    & 2\epsilon \cos{\phi} & [2\epsilon \sin{\phi} \cos{2\alpha} + 
    (\Delta G/2) \sin{2\alpha}] \\
    \Delta G/2 & \cos{2\alpha} & 0 & \sin{2\alpha} \\
    2\epsilon \cos{(\phi + \psi)} & \sin{2\alpha} \sin{\psi} 
    & \cos{\psi} & -\cos{2\alpha} \sin{\psi} \\
    2\epsilon \sin{(\phi + \psi)} & -\sin{2\alpha} \cos{\psi} & \sin{\psi} 
    & \cos{2\alpha} \cos{\psi}
\end{bmatrix},
\end{align}
where

\begin{enumerate}
\item $\Delta G$ is the relative calibration gain between the $X$ and $Y$ 
components of the feed, \\
\item $\psi$ is the phase difference between the the calibration ND signal 
and the incoming radiation, \\
\item $\alpha$ is a measure of the voltage ratio of the polarization 
ellipse produced when the feed observes pure linear polarization, \\
\item $\epsilon$ is a measure of imperfection of the feed in producing 
nonorthogonal polarizations (false correlations) in the two correlated 
outputs (represents undesirable cross coupling between the two polarizations), 
and \\
\item $\phi$ is the phase angle at which the voltage coupling $\epsilon$ 
occurs. It works with $\epsilon$ to couple $I$ with $(Q, U, V)$.
\end{enumerate}

Noting that for the linearly polarized calibration sources $V_\textrm{src} = 0$, 
and assuming that $I_\textrm{obs} = I_\textrm{src}$, dividing equation (\ref{eqn5}) 
by $I_\textrm{obs}$, and multiplying out the components gives 
\begin{eqnarray}
1 & = & 1 + [-2 \epsilon \sin{\phi} \sin{2 \alpha} + (\Delta G/2) \cos{2 \alpha}] 
\, (Q/I)_\text{src,rot} \; + \; 2\epsilon \cos{\phi} \, (U/I)_\text{src,rot} \label{eqn9}\\[5pt]
(Q/I)_\text{obs} & = & \Delta G/2 + \cos{2\alpha} \, (Q/I)_\text{src,rot} \label{eqn10}\\[5pt]
(U/I)_\text{obs} & = & 2\epsilon \cos{(\phi + \psi)} + \sin{2\alpha} \sin{\psi} 
\, (Q/I)_\text{src,rot} + \cos{\psi} \, (U/I)_\text{src,rot} \label{eqn11}\\[5pt]
(V/I)_\text{obs} & = & 2\epsilon \sin{(\phi + \psi)} - \sin{2\alpha} \cos{\psi} 
\, (Q/I)_\text{src,rot} + \sin{\psi} \, (U/I)_\text{src,rot} \ . \label{eqn12} 
\end{eqnarray}

\newpage
\section{Lo-Cal ND Mueller matrices}
\begin{table} [h]
\caption{GBT Mueller matrices for evenly spaced C-Band frequencies, calculated 
from two observations of 3C138 and one observation of B0529+075 using the Lo-Cal 
ND on 2021 November 27.}
\label{table:MM_Lo-Cal1} 
\vspace{2mm}
\begin{minipage}[c]{0.5\textwidth}
	\centering
	\textbf{4.300 GHz -- Lo-Cal ND} \\ [1.0ex] $ \left[
	\begin{array}{rrrr}
	   1.0000 &  0.0443 & -0.0150 &  0.0015 \\
         -0.0443 & -0.9977 &  0.0000 & -0.0678 \\
          0.0151 &  0.0059 & -0.9962 & -0.0865 \\
         -0.0002 & -0.0675 & -0.0867 &  0.9939 \\
	\end{array} \right]$
\end{minipage}
\begin{minipage}[c]{0.5\textwidth}
	\centering
	\textbf{5.900 GHz -- Lo-Cal ND} \\ [1.0ex] $ \left[
	\begin{array}{rrrr}
    	1.0000 & -0.0404 & -0.0070 &  0.0004 \\
        0.0404 & -1.0000 &  0.0000 &  0.0081 \\
        0.0070 & -0.0003 & -0.9992 & -0.0399 \\
        0.0003 &  0.0081 & -0.0399 &  0.9992 \\
	\end{array} \right]$
\end{minipage}
\\
\\ [4.0ex]
\begin{minipage}[c]{0.5\textwidth}
	\centering
	\textbf{4.700 GHz -- Lo-Cal ND} \\ [1.0ex] $ \left[
	\begin{array}{rrrr} 
		1.0000 & -0.0154 & -0.0039 &  0.0015 \\			
        0.0154 & -1.0000 &  0.0000 & -0.0081 \\			
        0.0038 &  0.0006 & -0.9974 & -0.0715 \\			
        0.0019 & -0.0081 & -0.0715 &  0.9974 \\			
	\end{array} \right]$
\end{minipage}
\begin{minipage}[c]{0.5\textwidth}
	\centering
	\textbf{6.300 GHz -- Lo-Cal ND} \\ [1.0ex] $ \left[
	\begin{array}{rrrr} 
		 1.0000 & 0.0194 & -0.0150 & -0.0096 \\
        -0.0198 & -0.9991 &  0.0000 &  0.0433 \\
         0.0155 & -0.0030 & -0.9976 & -0.0688 \\
        -0.0077 &  0.0432 & -0.0688 &  0.9967 \\
	\end{array} \right] $
\end{minipage}
\\ 
\\ [4.0ex]
\begin{minipage}[c]{0.5\textwidth}
	\centering
	\textbf{5.100 GHz -- Lo-Cal ND} \\ [1.0ex] $ \left[
	\begin{array}{rrrr}
	   1.0000 & -0.0065 & -0.0046 & -0.0023 \\
         0.0065 & -0.9999 &  0.0000 &  0.0147 \\
         0.0047 & -0.0011 & -0.9972 & -0.0748 \\
        -0.0020 &  0.0147 & -0.0748 &  0.9971 \\ 
	\end{array} \right] $
\end{minipage}
\begin{minipage}[c]{0.5\textwidth}
	\centering
	\textbf{6.700 GHz -- Lo-Cal ND} \\ [1.0ex] $ \left[
	\begin{array}{rrrr}
         1.0000 & -0.0232 &  0.0088 &  0.0164 \\
         0.0231 & -1.0000 &  0.0000 & -0.0059 \\
        -0.0094 &  0.0002 & -0.9993 & -0.0364 \\
         0.0162 & -0.0059 & -0.0364 &  0.9993 \\
	\end{array} \right] $
\end{minipage}
\\ 
\\ [4.0ex]
\begin{minipage}[c]{0.5\textwidth}
	\centering
	\textbf{5.500 GHz -- Lo-Cal ND} \\ [1.0ex] $ \left[
	\begin{array}{rrrr}
		 1.0000 & 0.0058 & -0.0124 & -0.0015 \\
        -0.0058 & -0.9999 &  0.0000 &  0.0143 \\
         0.0124 & -0.0011 & -0.9972 & -0.0749 \\
        -0.0005 &  0.0143 & -0.0749 &  0.9971 \\
	\end{array} \right] $
\end{minipage}
\begin{minipage}[c]{0.5\textwidth}
	\centering
	\textbf{7.100 GHz -- Lo-Cal ND} \\ [1.0ex] $ \left[
	\begin{array}{rrrr}
         1.0000 & -0.0362 & -0.0077 & -0.0110 \\
         0.0358 & -0.9994 &  0.0000 &  0.0341 \\
         0.0083 & -0.0017 & -0.9987 & -0.0509 \\
        -0.0118 &  0.0341 & -0.0510 &  0.9981 \\
	\end{array} \right] $
\end{minipage}
\\ [0.1ex]
\end{table}

\newpage
\begin{table} [ht]
\caption{GBT Mueller matrices for specific maser frequencies, calculated 
using Lo-Cal ND observations of 3C138 and B0529$+$075 on 2021 January 05, 
two observations of 3C138 and one observation of B0529+075 on 2021 
November 27. Data quality issues resulted in poor fitting for the 6.033 
and 6.181 GHz matrices so these results should be used with caution.}
\label{table:MM_Lo-Cal_maser} 
\vspace{2mm}
\begin{minipage}[c]{0.5\textwidth}
	\centering
	\textbf{4.765 GHz -- Lo-Cal ND} \\ [1.0ex] $ \left[
	\begin{array}{rrrr}
		 1.0000 & -0.0157 & -0.0036 & -0.0010 \\
         0.0157 & -1.0000 &  0.0000 &  0.0028 \\
         0.0037 & -0.0003 & -0.9959 & -0.0905 \\
        -0.0007 &  0.0028 & -0.0905 &  0.9959 \\
	\end{array} \right] $
\end{minipage}
\begin{minipage}[c]{0.5\textwidth}
	\centering
	\textbf{6.033 GHz -- Lo-cal ND} \\ [1.0ex] $ \left[
	\begin{array}{rrrr}
         1.0000 & -0.0496 & -0.0140 & -0.0200 \\
         0.0496 & -1.0000 &  0.0000 &  0.0000 \\
         0.0152 &  0.0000 & -0.9979 & -0.0645 \\
        -0.0191 &  0.0000 & -0.0645 &  0.9979 \\
 \end{array} \right] $
\end{minipage}
\\
\\ [4.0ex]
\begin{minipage}[c]{0.5\textwidth}
	\centering
	\textbf{4.751 GHz -- Lo-Cal ND} \\ [1.0ex] $ \left[
	\begin{array}{rrrr} 
         1.0000 & -0.0195 & -0.0016 & -0.0021 \\
         0.0195 & -1.0000 &  0.0000 & -0.0079 \\
         0.0017 &  0.0006 & -0.9967 & -0.0813 \\
        -0.0018 & -0.0079 & -0.0813 &  0.9967 \\ 
	\end{array} \right] $
\end{minipage}
\begin{minipage}[c]{0.5\textwidth}
	\centering
	\textbf{6.049 GHz -- Lo-Cal ND} \\ [1.0ex] $ \left[
	\begin{array}{rrrr}
         1.0000 & -0.0331 & -0.0089 & -0.0111 \\
         0.0330 & -1.0000 &  0.0000 &  0.0048 \\
         0.0093 & -0.0002 & -0.9993 & -0.0362 \\
        -0.0109 &  0.0048 & -0.0362 &  0.9993 \\
    \end{array} \right] $
\end{minipage}
\\ 
\\ [4.0ex]
\begin{minipage}[c]{0.5\textwidth}
	\centering
	\textbf{4.660 GHz -- Lo-Cal ND} \\ [1.0ex] $  \left[
	\begin{array}{rrrr}
         1.0000 & -0.0255 &  0.0142 & -0.0015 \\
         0.0255 & -1.0000 &  0.0000 & -0.0032 \\
        -0.0141 &  0.0003 & -0.9963 & -0.0854 \\
        -0.0026 & -0.0032 & -0.0854 &  0.9963 \\
    \end{array} \right] $
\end{minipage}
\begin{minipage}[c]{0.5\textwidth}
	\centering
	\textbf{6.181 GHz -- Lo-Cal ND} \\ [1.0ex] $  \left[
	\begin{array}{rrrr}
         1.0000 & -0.0427 & -0.0037 & -0.0019 \\
         0.0427 & -0.9999 &  0.0000 & -0.0117 \\
         0.0039 &  0.0016 & -0.9908 & -0.1357 \\
        -0.0009 & -0.0116 & -0.1357 &  0.9907 \\
    \end{array} \right] $
\end{minipage}
\\ 
\\ [4.0ex]
\begin{minipage}[c]{0.5\textwidth}
	\centering
	\textbf{4.830 GHz -- Lo-Cal ND} \\ [1.0ex] $  \left[
	\begin{array}{rrrr}
         1.0000 & -0.0038 & -0.0028 & -0.0004 \\
         0.0038 & -1.0000 &  0.0000 &  0.0099 \\
         0.0028 & -0.0008 & -0.9967 & -0.0811 \\
        -0.0002 &  0.0098 & -0.0811 &  0.9967 \\
    \end{array} \right] $
\end{minipage}
\begin{minipage}[c]{0.5\textwidth}
	\centering
	\textbf{6.668 GHz -- Lo-Cal ND} \\ [1.0ex] $  \left[
	\begin{array}{rrrr}
         1.0000 &  0.0070 & -0.0021 &  0.0035 \\
        -0.0070 & -1.0000 &  0.0000 &  0.0036 \\
         0.0019 & -0.0003 & -0.9976 & -0.0696 \\
         0.0036 &  0.0036 & -0.0696 &  0.9976 \\
	\end{array} \right] $
\end{minipage}
\\ [0.1ex]
\end{table}

\begin{table}  [h]
\caption{GBT Mueller matrices at 4.700 GHz for observations with different 
calibration noise diodes. Differences are apparent, in particular, in the 
$m_{IQ}$ and $m_{QI}$ terms which represent the relative calibration gain 
in the $X$ and $Y$ channels.}
\label{table:MM-Lo_vs_Hi}
\vspace{2mm}
\begin{minipage}[c]{0.5\textwidth}
	\centering
	\textbf{4.700 GHz -- Hi-Cal ND} \\ [1.0ex] $ \left[
	\begin{array}{rrrr}
         1.0000 & -0.0009 &  0.0020 & -0.0007 \\
         0.0009 & -1.0000 &  0.0000 & -0.0004 \\
        -0.0019 &  0.0000 & -0.9946 & -0.1042 \\
        -0.0009 & -0.0004 & -0.1042 &  0.9946 \\
	\end{array} \right] $
\end{minipage}
\begin{minipage}[c]{0.5\textwidth}
	\centering
	\textbf{4.700 GHz -- Lo-Cal ND} \\ [1.0ex] $\left[
	\begin{array}{rrrr} 
		1.0000 & -0.0154 & -0.0039 & 0.0015 \\
        0.0154 & -1.0000 & 0.0000 & -0.0081 \\
        0.0038 & 0.0006 & -0.9974 & -0.0715 \\
        0.0019 & -0.0081 & -0.0715 & 0.9974 \\
     \end{array} \right]$
\end{minipage}
\\ [0.1ex]
\end{table}


\begin{thebibliography}{99}

\bibitem[Fallon(2022)]{Fallon2022} Fallon, P. 2022, GBT Memo 306: Frequency-switching 
and position-switching GBTIDL analysis code used for determining Stokes and polarization 
parameters

\bibitem[Goddy et al.(2020)]{Goddy_et_al_2020} Goddy, J., Masters, K. \& Stark, D.\
2020, American Astronomical Society Meeting Abstracts \#235, 167

\bibitem[Heiles et al.(2001)]{Heiles_et_al_2001} Heiles, C., Perillat, P., Nolan, M., 
et al.\ 2001, \pasp, 113, 1274. doi:10.1086/323289

\bibitem[Heiles(2002)]{Heiles2002} Heiles, C.\ 2002, Single-Dish Radio Astronomy: 
Techniques and Applications, 278, 131

\bibitem[Heiles et al.(2003)]{Heiles_et_al_2003} Heiles, C., Robishaw, T., Troland, 
T, \& Roshi, D.-A.\ 2003, GBT Memo 23: Calibrating the GBT at L, C, and X Bands

\bibitem[Heiles \& Robishaw(2022)]{HR2022} Heiles, C. \& Robishaw, T.\ 2022, 
All-Stokes Single Dish Data with the RHSTK\_2021 Software Package, Unpublished

\bibitem[IEEE(2018)]{IEEE18} IEEE\ 2018, IEEE Standard Definitions of Terms for Radio Wave Propagation, IEEE Std 211-2018 (Revision of IEEE Std 211-1997), pp.1-57, 1 Feb. 2019, doi: 10.1109/IEEESTD.2019.8657413.

\bibitem[Liao et al.(2016)]{Liao_et_al_2016} Liao, Y.-W., Chang, T.-C., Kuo, C.-Y., 
et al.\ 2016, \apj, 833, 289. doi:10.3847/1538-4357/833/2/289

\bibitem[Prestage et al.(2015)]{Prestage_2015} Prestage, R.~M., Bloss, M., Brandt, J., et al.\ 2015, \\ 2015 URSI-USNC Radio Science Meeting, 4. doi:10.1109/USNC-URSI.2015.7303578

\bibitem[Robishaw \& Heiles(2006)]{RH2006} Robishaw, T. \& Heiles, C.\ 2006, GBT 
Memo 244: The Proper Production of Full-Stokes Spectra Using The Fully-Functioning 
Full-Stokes Mode of the GBT Spectrometer

\bibitem[Robishaw \& Heiles(2021)]{RH2021} Robishaw, T. \& Heiles, C.\ 2021, The 
WSPC Handbook of Astronomical Instrumentation, 127. New Jersey: World Scientific.

\end{thebibliography}
\end{document}